\documentclass[USenglish,a4paper]{lipics-v2021}
\nolinenumbers
\usepackage{amsthm, amssymb, enumerate, mathtools}
\usepackage{interval,braket}
\intervalconfig{scaled,soft open fences}
\usepackage[alignedforall]{lpform}
\usepackage{subcaption}
\usepackage{amsfonts}
\usepackage{booktabs}
\usepackage{multirow}
\usepackage{multicol}
\usepackage{csquotes}
\usepackage{phoenician}

\usepackage{stmaryrd}

\usepackage{xcolor}
\usepackage{algorithm}
\usepackage{algpseudocode}

\usepackage{adjustbox}

\usepackage{pgfplots,tikz}

\makeatletter

\tikzset{drawscheduleapplystyle/.code={\tikzset{#1}}}

\def\schedule@draw<#1>[#2] at(#3) #4;{
  \edef\globalopts{#2}
  \if\relax\detokenize{#1}\relax
    \let\schedule@list\empty
    \foreach \x in {#4} {
      \ifx\schedule@list\empty
        \xdef\schedule@list{\x}
      \else
        \xdef\schedule@list{\x,\schedule@list}
      \fi
    }
  \else
    \xdef\schedule@list{#4}
  \fi
  \xdef\drawscheduleWidth{0.0}
  \begin{scope}[shift={(#3)}]
  \foreach [count=\i from 0] \machine/\w/\machineopts in \schedule@list {
    \edef\newWidth{\drawscheduleWidth+\w}
    \xdef\drawscheduleTime{0.0}
    \ifdim \w pt > 0 pt
      \foreach \t/\labl/\jobopts/\nodeopts in \machine {
        \edef\newTime{\drawscheduleTime+\t}
        \if\relax\detokenize{#1}\relax
          \draw [drawscheduleapplystyle/.expand once=\globalopts,drawscheduleapplystyle/.expand once=\machineopts,drawscheduleapplystyle/.expand once=\jobopts] (\drawscheduleTime,\drawscheduleWidth) rectangle (\newTime,\newWidth) node[pos=.5,drawscheduleapplystyle/.expand once=\nodeopts] {\labl};
        \else
          \draw [drawscheduleapplystyle/.expand once=\globalopts,drawscheduleapplystyle/.expand once=\machineopts,drawscheduleapplystyle/.expand once=\jobopts] (\drawscheduleWidth,\drawscheduleTime) rectangle (\newWidth,\newTime) node[pos=.5,drawscheduleapplystyle/.expand once=\nodeopts] {\labl};
        \fi
        \xdef\drawscheduleTime{\newTime}
      }
      \xdef\drawscheduleWidth{\newWidth}
    \else
      \xdef\drawscheduleWidth{\drawscheduleWidth-\w}
    \fi
  }
  \end{scope}
}

\newcommand\schedule{}
\def\schedule{\schedule@checkorient}
\def\schedule@checkorient{\pgfutil@ifnextchar u{\schedule@up}{\schedule@checkops<>}}
\def\schedule@up up#1;{\schedule@checkops<u>#1;}
\def\schedule@checkops<#1>{\pgfutil@ifnextchar[{\schedule@ops<#1>}{\schedule@checkat<#1>[]}}
\def\schedule@checkat<#1>[#2]{\pgfutil@ifnextchar a{\schedule@at<#1>[#2]}{\schedule@draw<#1>[#2] at(0,0) }}
\def\schedule@ops<#1>[#2]#3;{\schedule@checkat<#1>[#2]#3;}
\def\schedule@at<#1>[#2]#3at#4(#5) #6;{\schedule@draw<#1>[#2] at(#5) #6;}

\newcommand\schedule@drawlines[4][]{
  \foreach \t/\labl/\drawopts/\nodeopts in {#4} {
    \if\relax\detokenize{#1}\relax
      \draw [dash pattern={on 1.5pt off 0.8pt},drawscheduleapplystyle/.expand once=\drawopts] (\t,#3+#2) -- (\t,-#2) node [below,drawscheduleapplystyle/.expand once=\nodeopts] {\labl};
    \else
      \draw [dash pattern={on 1.5pt off 0.8pt},drawscheduleapplystyle/.expand once=\drawopts] (-#2,\t) -- (#3+#2,\t) node [right,drawscheduleapplystyle/.expand once=\nodeopts] {\labl};
    \fi
  }
}

\newcommand\vlines[3][.2]{\schedule@drawlines{#1}{#2}{#3}}
\newcommand\hlines[3][.2]{\schedule@drawlines[h]{#1}{#2}{#3}}

\makeatother \usetikzlibrary{math,shapes,arrows,fit,calc,positioning,patterns,decorations.pathmorphing,decorations.pathreplacing,calligraphy,patterns,arrows.meta,external}
\tikzset{ brokenrect/.style={

    append after command={

      \pgfextra{

      \path[draw,#1]

       decorate[decoration={zigzag,segment length=0.3em, amplitude=.7mm}]

       {(\tikzlastnode.north east)--(\tikzlastnode.south east)}      

        -- (\tikzlastnode.south west)|-cycle;

        }}}}

\tikzset{ brokenrect2/.style={

    append after command={

      \pgfextra{

      \path[draw,#1]

       decorate[decoration={zigzag,segment length=0.3em, amplitude=.7mm}]

       {(\tikzlastnode.north west)--(\tikzlastnode.south west)}      

        -- (\tikzlastnode.south east)|-cycle;

        }}}}
        
\tikzset{cross/.style={cross out, draw=black, minimum size=2*(#1-\pgflinewidth), inner sep=0pt, outer sep=0pt},
cross/.default={1pt}}

\usepackage{etoolbox}
\newcommand{\threefield}[3]{\({\normalfont\mathrm{#1}\allowbreak\mathbf{\vert}\allowbreak\mathrm{#2}\allowbreak\mathbf{\vert}\allowbreak\mathrm{#3}}\)}%

\newcommand{\Cmax}{C_{\max}}
\newcommand{\Cmin}{C_{\min}}

\newcommand\PCmax[1][]{\threefield P {#1} {\Cmax}\xspace}

\newcommand\smax{s_{\max}}

\providecommand\super[2]{#2^{(#1)}}

\newcommand\HM[1][m,n]{\ifmmode\mathit{HM}(#1)\else\ensuremath{\HM[#1]}\xspace\fi}
\newcommand\HMn{\ifmmode\mathit{HM}(n)\else\ensuremath{\HM[n]}\xspace\fi}

\RequirePackage{interval}
\intervalconfig{separator symbol = {;}}

\newcommand\iv[1]{[{#1}]}

\makeatletter
\RequirePackage{mathtools}
\newcommand\DeclareMathOperators[1]{%
  \@for\@ii:=#1\do{\expandafter\DeclareMathOperator\@ii}%
}%
\makeatother
\DeclareMathOperators{
  \Oh{\mathcal{O}},\oh{o},
  \OPT{OPT},\chp{chp},\load{L},
  \ks{ks},\cks{cks},
  \size{size},\pop{pop},\push{push},\parent{parent},
  \lcm{lcm},\supp{supp},\sign{sign},\poly{poly}
}

\newcommand{\op}[1]{\operatorname{#1}}%

\DeclarePairedDelimiter\abs{\lvert}{\rvert}%
\DeclarePairedDelimiter\ceil{\lceil}{\rceil}
\DeclarePairedDelimiter\floor{\lfloor}{\rfloor}

\renewcommand{\emptyset}{\varnothing}%
\newcommand{\eps}{\varepsilon}%
\newcommand{\Z}{\mathbb{Z}}%
\newcommand{\N}{\mathbb{N}_{\geq 0}}%
\newcommand{\Q}{\mathbb{Q}}%
\newcommand{\R}{\mathbb{R}}%
\newcommand{\NP}{\mathcal{NP}}%

\newcommand\dotcup{\mathbin{\dot\cup}}
\providecommand\ucup\dotcup
\newcommand{\cupdot}{\mathbin{\dot{\cup}}}%
\makeatletter
\@ifclassloaded{acmart}{}{\newcommand{\bigcupdot}{\mathop{\dot{\bigcup}}}}%
\makeatother

\newcommand\defeq\coloneqq
\newcommand\eqdef\eqqcolon

\RequirePackage{dsfont}
\NewDocumentCommand\onevec{o}{%
	\IfValueTF{#1}{%
		\mathds 1_{#1}%
	}{%
		\mathds 1%
	}%
}
\NewDocumentCommand\zerovec{o}{%
	\IfValueTF{#1}{%
		\mathds O_{#1}%
	}{%
		\mathds O_{\N \times 1}%
	}%
}
\NewDocumentCommand\zero{o}{%
	\IfValueTF{#1}{\mathds O_{#1}}{\mathds O}%
}
\NewDocumentCommand\one{o}{%
	\IfValueTF{#1}{\mathds E_{#1}}{\mathds E}%
}

\usepackage{bbm}
\usepackage{bm}
\NewDocumentCommand\uvec{om}{%
	\IfValueTF{#1}{%
		\mathbbm e^{(#2)}_#1
	}{%
		\mathbbm e^{(#2)}
	}%
}

\newcommand{\ie}{i.\,e.\xspace}

\newtheorem{property}{Property}

\usepackage{xfrac}

\newcommand{\PCmaxno}{\PCmax[setup=s_i]}
\newcommand{\PCmaxpm}{\PCmax[pmnt,setup=s_i]}
\newcommand{\PCmaxsp}{\PCmax[split,setup=s_i]}
\newcommand{\OPTpm}{\OPT}

\newcommand\mc\mathcal
\newcommand{\encircle}[1]{\textnormal{\small\textcircled{\scriptsize{#1}}}}
\newcommand{\encircleinsched}{{circle,baseline=-0.5ex,inner sep=.05cm,draw=black,fill=white}}

\newcommand{\med}{\ensuremath{\op{med}}}
\newcommand{\wrap}{\textsc{Wrap}\xspace}
\newcommand{\wrapdyn}{\textsc{WrapLJ}\xspace}

\newcommand{\typeone}{\encircle{1}\xspace}
\newcommand{\typetwo}{\encircle{2}\xspace}
\newcommand{\typethree}{\encircle{3}\xspace}
\newcommand{\typefour}{\encircle{4}\xspace}
\newcommand{\typefive}{\encircle{5}\xspace}
\newcommand{\typesix}{\encircle{6}\xspace}
\newcommand{\typeseven}{\encircle{7}\xspace}
\newcommand{\typeeight}{\encircle{8}\xspace}
\newcommand{\typenine}{\encircle{9}\xspace}
\newcommand{\typeten}{\encircle{10}\xspace}

\newcommand{\mnice}{m_{\op{nice}}}
\newcommand{\Mnice}{M_{\op{nice}}}
\newcommand{\Msingle}{M_{\op{single}}}
\newcommand{\Mdual}{M_{\op{dual}}}
\newcommand{\Mnine}{M_{9}}
\newcommand{\Mten}{M_{10}}
\newcommand{\Mnt}{M_{9,10}}
\newcommand{\Met}{M_{8,10}}
\newcommand{\Mrem}{M_{\op{rem}}}
\newcommand{\Snt}{S_{9,10}}
\newcommand{\Mlarge}{M_{\op{large}}}

\newcommand{\Inice}{I_{\op{nice}}}
\newcommand{\Snice}{S_{\op{nice}}}
\newcommand{\Lnice}{L_{\op{nice}}}
\newcommand{\Srest}{S_{\op{rest}}}
\newcommand{\Lrest}{L_{\op{rest}}}
\newcommand{\Ieightrem}{I_{8 \op{rem}}}
\newcommand{\Seight}{S_{8,10}}
\newcommand{\Meight}{M_{8,10}}
\newcommand{\Sseven}{S_{\typeseven}}

\newcommand{\ieight}{i_{8}}
\newcommand{\inine}{i_{9}}
\newcommand{\iten}{i_{10}}
\newcommand{\inineten}{C_{9,10}}
\newcommand{\Pext}{P^{\op{ext}}}

\newcommand{\Pover}{P^{\op{over}}}
\newcommand{\Jone}{\mc{J}_1}
\newcommand{\Jtwo}{\mc{J}_2}

\newcommand{\solstr}{\sigma^{\sfrac43}}
\newcommand{\solopt}{\sigma^{\OPT}}

\newcommand\tsuper[2]{\super {\text{\normalfont #1}} #2}
\providecommand\tred{\mathrm{red}}
\providecommand\tblue{\mathrm{blue}}
\providecommand\textcite[1]{\citet{#1}}

\NewDocumentCommand{\lexobj}{s}{%
	{%
		\tsuper {\IfBooleanTF#1{mod-lex}{lex}}	\varphi%
	}%
}

\newcommand\lb\ell\newcommand\mb\Lambda

\providecommand\nub[1][\relax]{\super \tblue {\expandafter #1 \nu}}
\providecommand\mub[1][\relax]{\super \tblue {\expandafter #1 \mu}}
\providecommand\mur[1][\relax]{\super \tred {\expandafter #1 \mu}}

\newcommand\thecrit{\operatorname{critical-type}}
\NewDocumentCommand\crit{s O{} d()}{
	\IfNoValueTF{#3}{
		\dot t
	}{
		\thecrit
		\IfNoValueTF{#1}{
			\enpar[#2]{\allowbreak #3}
		}{
			\enpar*[#2]{\allowbreak #3}
		}
	}
}

\providecommand\crit{\operatorname{critical-type}}

\let\OPT=\relax
\NewDocumentCommand\OPT{t* t+ t-}{
	\let\suffix=\relax
	\IfBooleanT#1{
		\newcommand\suffix{^\star}
	}
	\IfBooleanT#2{
		\newcommand\suffix{^{\uparrow}}
	}
	\IfBooleanT#3{
		\newcommand\suffix{^{\downarrow}}
	}
	\operatorname{OPT\suffix}
}
\NewDocumentCommand\newlname{m m}{
	\IfNoValueTF{#1}{
		,name=\texttt{#2}
	}{
		,name={#1\texttt{[#2]}}
	}
}
\NewDocumentCommand\lname{s m}{%
	\IfBooleanTF#1{%
		{~\texttt{[#2]}}%
	}{%
		\texttt{#2}%
	}%
}

\newcommand\nfoldN{N}
\newcommand\nfoldR{R}
\newcommand\nfoldS{S}
\newcommand\nfoldT{H}
\newcommand\nfoldn\nfoldN\newcommand\nfoldr\nfoldR\newcommand\nfolds\nfoldS\newcommand\nfoldt\nfoldT

\DeclarePairedDelimiter{\internalXenpar}{\lparen}{\rparen}
\NewDocumentCommand{\enpar}{s O{} t~ m}{
	\IfBooleanTF{#3}{%
		{%
		\IfBooleanTF{#1}{\internalXenpar*{#4}}{\internalXenpar[#2]{#4}}%
		}%
	}{%
		\IfBooleanTF{#1}{\internalXenpar*{#4}}{\internalXenpar[#2]{#4}}%
	}%
}

\newlength{\IrresponsibleFantasy}

\newcommand\aref[1]{\hyperref[#1]{\autoref*{#1} (\nameref*{#1})}}
\usepackage{lineno}
\let\lineref=\relax
\newcommand\lineref[1]{\hyperref[#1]{Line~\ref*{#1}}}
\newcommand\stepref[1]{\hyperref[#1]{Step~\ref*{#1}}}
\newcommand\ineqref[1]{\hyperref[#1]{Inequality~\ref*{#1}}}
\newcommand\constraintref[1]{\hyperref[#1]{Constraint~\ref*{#1}}}
\newcommand\caseref[2]{\hyperref[#1]{Case~#2}}

\NewDocumentCommand\case{o m}{\noindent \underline{\IfNoValueF{#1}{Case #1:\xspace}#2}\xspace}
	\makeatletter
	\newcommand{\dotminus}{\mathbin{\text{\@dotminus}}}
	\newcommand{\@dotminus}{%
		\ooalign{\hidewidth\raise1ex\hbox{.}\hidewidth\cr$\m@th-$\cr}%
	}
	\makeatother
{
\let\olddotminus=\dotminus
\renewcommand\dotminus{\mathrel{\olddotminus}}
}

 \usepackage{nameref}
\usepackage{scrextend}
\usepackage{xspace}
\title{A $(4/3+\varepsilon)$-Approximation for Preemptive Scheduling with Batch Setup Times}

\author{Max A. Deppert}{Kiel University, Kiel, Germany}{made@informatik.uni-kiel.de}{https://orcid.org/0000-0003-3083-7998}{}
\author{David Fischer}{Helmut Schmidt University, Hamburg, Germany}{fischerd@hsu-hh.de}{https://orcid.org/0000-0001-8402-1818}{}
\author{Klaus Jansen}{Kiel University, Kiel, Germany}{kj@informatik.uni-kiel.de}{https://orcid.org/0000-0001-8358-6796}{}
\authorrunning{M.A.\@ Deppert \and D.\@ Fischer \and K.\@ Jansen}


\begin{document}

\maketitle

\begin{abstract}
We consider the $\mathcal{NP}$-hard problem $\mathrm{P} \mathbf{\vert} \mathrm{pmtn, setup=s_i} \mathbf{\vert} \mathrm{C_{\max}}$, the problem of scheduling $n$ jobs, which are divided into $c$ classes, on $m$ identical parallel machines while allowing preemption.
For each class $i$ of the $c$ classes, we are given a setup time $s_i$ that is required to be scheduled whenever a machine switches from processing a job of one class to a job from another class.
The goal is to find a schedule that minimizes the makespan.

We give a $(4/3+\varepsilon)$-approximate algorithm with run time in $\mathcal{O}(n^2 \log(1/\varepsilon))$.
For any $\varepsilon < 1/6$, this improves upon the previously best known approximation ratio of $3/2$ for this problem.

Our main technical contributions are as follows.
We first partition any instance into an \enquote{easy} and a \enquote{hard} part, such that a $4/3 T$-approximation for the former is easy to compute for some given makespan $T$.
We then proceed to show our main structural result, namely that there always exists a $4/3 T$-approximation for any instance that has a solution with makespan $T$, where the hard part has some easy to compute properties.
Finally, we obtain an algorithm that computes a $(4/3+\varepsilon)$-approximation in time n $\mathcal{O}(n^2 \log(1/\varepsilon))$ for general instances by computing solutions with the previously shown structural properties.
\end{abstract}

\thispagestyle{empty}\clearpage
\setcounter{page}{1}

\section{Introduction}
\label{sec:intro}

In this work, we investigate the scheduling problem with setup times on identical parallel machines allowing preemption, with the objective of minimizing the makespan.
This problem is denoted as \PCmaxpm in Graham-Notation~\cite{graham79}.
Formally, an instance of \PCmaxpm consists of $m \in \N$ identical and parallel machines, a set $J$ of $n \in \N$ jobs $j \in J$, a partition $\bigcupdot_{i=1}^c C_i = J$ of $J$ into $c$ nonempty and disjoint subsets $C_i \subseteq J$, called classes, a processing time $p_j \in \N$ for each jobs $j \in J$, and a setup time of $s_i \in \N$ for each class $i$.
Allowing preemption means that each job $j$ may be split into any number of job pieces $j_1, \dots, j_d$, with $p_{j_1}, \dots, p_{j_d} \in \R$ and $\sum_{k = 1}^d p_{j_k} = p_j$, but no job pieces $j_k, j_{k'}$ of the same job $j$ may be scheduled at the same time over all machines.
We also say that a job $j$ may not be \emph{parallelized}.
A job $j$ is (fully) scheduled iff all of its job pieces are scheduled. 

The objective of \PCmaxpm is to find a feasible schedule $\sigma$ which minimizes the makespan, i.e. the maximum completion time of any machine in $\sigma$.
A schedule $\sigma$ is feasible for an instance $I$ of \PCmaxpm iff all jobs are fully scheduled on at most $m$ machines, on each machine $q \in \iv m$, before any number of job pieces of the same class $i \in \iv c$ are scheduled, there is a setup time $s_i$ scheduled, and no job is parallelized.
\PCmaxpm is $\NP$-hard already for $m = 2$~\cite{MP89}.
We thus focus on giving approximate solutions for this problem in polynomial time.

\subsection{Related Work} The problem \PCmaxpm was first introduced by Monma and Potts~\cite{MP89}, where they not only proved the problems $\NP$-hardness, but also gave a dynamic programming approach computing exact solutions to this problem in $\Oh(n^c)$.
They were also the first to give a constant factor approximation algorithms for the problem~\cite{MP93}, one with general approximation ratio of $\sfrac{5}{3}$ and run time in $\Oh(n+(m+c)\log(m+c))$, \ie exponential run time for general instances, and another resembling the Wrap-Around-Rule introduced by McNaughton~\cite{M59}, with approximation ratio of $2 - \sfrac{1}{(\floor*{m/2} + 1)} = \lim_{m \to \infty} 2$ and run time in $\Oh(n)$.
Deppert and Jansen~\cite{DJ19} improved on this result, giving an algorithm for \PCmaxpm with approximation ratio of $\sfrac32$ and slightly worse run time in $\Oh(n \log(n))$.
 
Notable, related variations of this problem include the problem \PCmaxpm restricted to \emph{single-job-batches}, \ie $\abs{C_i} = 1$ for all $i \in \iv c$, the splittable version of the problem \PCmaxsp, where jobs may not only be preempted, but also parallelized, and the non-preemptive version \PCmaxno.
For \PCmaxpm with single-job-batches, Schuurman and Woeginger~\cite{SW99} gave a $(\sfrac43 + \eps)$-approximation, as well as a PTAS for the case of $s_i = s, \forall i \in \iv c$.
For \PCmaxsp, Xing and Zhang~\cite{XZ00} gave a $\sfrac53$-approximation algorithm.
For \PCmaxno, there exists a series of constant-factor approximation algorithms~\cite{JL16,MMMR15}, as well as a PTAS~\cite{JL16}.
Research on approximation algorithms for these three problem variants recently culminated when Jansen et al.~\cite{JKMR19} introduced EPTASs for each, based on a new $n$-fold integer program formulation and using an efficient algorithm for such integer programs~\cite{HOR13}.
For a concise history of approximating \PCmaxpm and related problems, we refer the reader to the overview given by Deppert and Jansen~\cite{DJ19}.
Interestingly however, there does not yet exist an (E)PTAS for general \PCmaxpm. Therefore, the $\sfrac32$-approximation algorithm by Deppert and Jansen~\cite{DJ19} gives the currently best known approximation for this problem.

\subsection{Our Contribution}
We give a $(\sfrac43+\eps)$-approximate algorithm for \PCmaxpm with run time in $\Oh(n^2 \log(1/\eps))$.
For any $\eps < \sfrac16$, this improves upon the previously best known approximation factor of $\sfrac32$ of a polynomial time algorithm by Deppert and Jansen~\cite{DJ19} for this problem.

\noindent\textbf{Organization of this work.}
We first give some preliminary notions in \autoref{sec:prelims}, stating the basic notation used throughout this work, presenting how we structure instances of the problem, and introducing the important concept of \enquote{Batch Wrapping}.
This is followed by \autoref{sec:technical_overview}, a rough technical overview of our core ideas.
Our main results are then presented in multiple stages: We first give a \(\sfrac43 T\)-dual approximation for \enquote{nice instances} in \autoref{sec:nice_instances}, followed by a structural analysis of general instances in \autoref{sec:structure}.
These ideas are then synthesized into the $(\sfrac43+\eps)$-approximate algorithm for general instances of \PCmaxpm in \autoref{sec:approxalgo}.
We finish by giving a short conclusion and an outlook on future research in \autoref{sec:concl}.

\section{Preliminaries}
\label{sec:prelims}
\noindent\textbf{Notation.}
We begin by giving some notation.
A \emph{job piece} $j'$ of a job $j$ is a new job $j'$ with processing time $p_{j'} \leq p_j$ of class $i$, iff $j$ is of class $i$.
A job $j$ is fully scheduled if all of its job pieces $j'$ are scheduled, and the total processing times of its job pieces is $p_j$, \ie $\sum_{j' \in J} p_{j'} = p_j$.
We denote the total processing time of all jobs of a class as $P(C_i)$, \ie $P(C_i) \defeq \sum_{j \in C_i} p_j$.
For readability purposes, we abuse the notation slightly and may use $s_j$ to refer to the setup time $s_i$ of class $i$ with $p_j \in C_i$.
We also may write $s_x$ or $C_x$ to refer to the setup time or job set of a class $i_x$, respectively.
The load $L_{\sigma}(q)$ of a machine $q$ is the sum of all setup times and all processing times of job pieces scheduled on $q$ in $\sigma$.
If $\sigma$ is clear from context, we just write $L(q)$.
The load $L(S)$ of a set of setup times and job pieces $S$ is the total sum over all these elements, \ie $L(S) := \sum_{s \in S} s$.
We say a class $i$ is \enquote{fully scheduled} on a machine $q$, w.r.t. some schedule $\sigma$, if all job (pieces) in $C_i$ are scheduled on $q$ in $\sigma$.
By $\lambda^{\sigma}_i$, we denote the number of setup times $s_i$ of class $i$ scheduled in $\sigma$. 

An instance $I' = (m',J,\bigcupdot_{i=k}^c C_k, p'_1, \dots, p'_{\abs{J}}, s_1, \dots, s_c)$ is a sub-instance of an instance $I = (m,J,\bigcupdot_{i=k}^c C_k,  p_1, \dots, p_{\abs{J}}, s_1, \dots, s_c)$ iff $m' \leq m$ and $p'_j \leq p_j, \forall j \in J$.
If $I'$ is a sub-instance of a (sub-)instance $I$, it can be defined solely by $m' \leq m$ and $p'_1, \dots, p'_{\abs{J}}$, as all other parameters are the same for both $I$ and $I'$.

Let $T$ be a guess on the optimal makespan of an instance $I$ of \PCmaxpm, and $\alpha_i \defeq \ceil{P(C_i)/(T-s_i)} \geq \floor{P(C_i)/(T-s_i)} =: \alpha_i'$.
\begin{lemma}
\label{lemma:lambda_geq_alpha}
For any schedule $\sigma$ with makespan $T$, $\lambda^{\sigma}_i \geq \alpha_i, \forall i \in \iv c$.
\end{lemma}
\begin{proof}
In any schedule $\sigma$ with makespan $T$, there is at most $(T-s_i)$ processing time of each class $i$ scheduled on a single machine.
Thus, there are at least $\ceil{P(C_i)/(T-s_i)} = \alpha_i$ different machines required to schedule the total processing time $P(C_i)$ of a class $i$ in a feasible schedule.
To be feasible, there must be at least one setup time $s_i$ scheduled on each of the $\alpha_i$ machines, implying $\lambda^{\sigma}_i \geq \alpha_i, \forall i \in \iv c$.
\end{proof}

\subsection{Structuring an Instance.}
\begin{figure}[h]
	\centering
	\begin{adjustbox}{width=\linewidth, keepaspectratio}
	{\tikz{%
\begin{axis}[
    axis lines=left,
    xmin = 0, xmax = 2.1,
    ymin = 0, ymax = 1.1,
    y=\textwidth/2.75,
    x=\textwidth/2.75,
    xtick={1/3,4/9,1/2,2/3,5/6,1,4/3,5/3},
    xticklabels={$\frac13$,$\frac49$,$\frac12$,$\frac23$,$\frac56$,1,$\frac43$,$\frac53$},
    ytick={1/3,1/2,2/3,1},
    yticklabels={$\frac13$,$\frac12$,$\frac23$,1},
    xlabel={$s_i + P(C_i)$},
    ylabel={$s_i$}
  ]
  \draw (axis cs:0,0) -- (axis cs:1,1);
  \draw (axis cs:4/3,2/3) -- (axis cs:5/3,1/3);

  \draw (axis cs:1,1) -- (axis cs:2,1);
  \draw[dashed] (axis cs:2,1) -- (axis cs:2.1,1);
  \draw (axis cs:1,2/3) -- (axis cs:2,2/3);
  \draw[dashed] (axis cs:2,2/3) -- (axis cs:2.1,2/3);
  \draw (axis cs:1/3,1/3) -- (axis cs:2,1/3);
  \draw[dashed] (axis cs:2,1/3) -- (axis cs:2.1,1/3);

  \draw (axis cs:4/9,1/3) -- (axis cs:4/9,4/9);
  \draw (axis cs:1/2,1/3) -- (axis cs:1/2,1/2);
  \draw (axis cs:2/3,1/3) -- (axis cs:2/3,2/3);
  \draw (axis cs:5/6,1/3) -- (axis cs:5/6,5/6);
  \draw (axis cs:1,1/3) -- (axis cs:1,1);
  \draw (axis cs:4/3,1/3) -- (axis cs:4/3,2/3);

  \node[] at (axis cs:1+5/6,5/6) {\typeone};
  \node[] at (axis cs:1+5/6,1/2) {\typetwo};
  \node[] at (axis cs:4/3+3/24,1/2-1/12) {\typethree};
  \node[] at (axis cs:1+1/6,1/2) {\typefour};
  \node[] at (axis cs:1/2+1/12,1/2-1/12) {\typefive};
  \node[] at (axis cs:29/72,13/36) {\typesix};
  \node[] at (axis cs:4/3+1/6,1/6) {\typeseven};
  \node[] at (axis cs:17/36,15/36) {\typeeight};
  \node[] at (axis cs:2/3+1/12,1/2) {\typenine};
  \node[] at (axis cs:5/6+1/12,1/2+1/12) {\typeten};
\end{axis} %
}}
	\end{adjustbox}
	\caption{A graphic representation of the total partition of the classes of an instance.}
	\label{fig:preemptive:class_partition}
\end{figure}
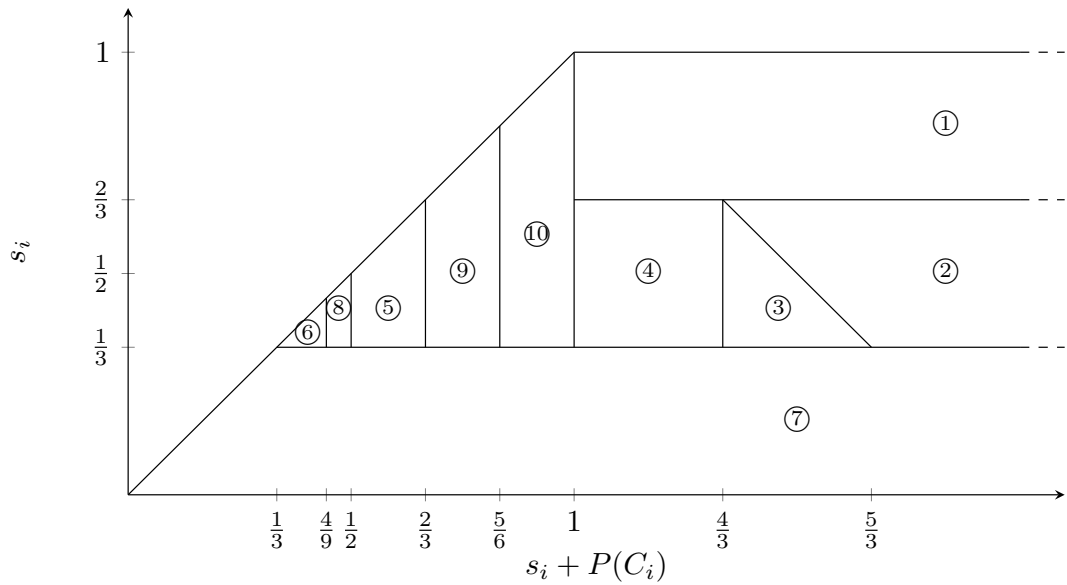

For some guess on the makespan $T$, we structure any instance of \PCmaxpm by partitioning the classes of an instance into sets $\typeone,\dots,\typeten \subseteq [c]$ such that $[c] = \typeone \cup \typetwo \cup \typethree \cup \typefour \cup \typefive \cup \typesix \cup \typeseven \cup \typeeight \cup \typenine \cup \typeten$, as defined in the \autoref{tab:preemptive:class_partition} and shown \autoref{fig:preemptive:class_partition}, which we call \enquote{types}.
For readability purposes, here, we assume that $T=1$, and set $\exp = [\sfrac23,1]$, $\med = [\sfrac13,\sfrac23)$, and $\chp = [0,\sfrac13)$.
This is a total partition of classes in any instance; each occurring class belongs to exactly one set from \typeone to \typeten.
This structuring will help us deal with the different types of classes: For classes of the same type, we will use similar algorithmic ideas.

\begin{table}[h]
    \centering
    \begin{tabular}{|c|c|c|}
        \hline
            Set & $s_i$ & $s_i + P(C_i)$\\\hline
            \typeone & $\exp$ & $\big[1,m\big]$\\
            \typetwo & $\med$ & $[2-s_i,m]$\\
            \typethree & $\med$ & $\big(\sfrac43,2-s_i\big)$\\
            \typefour & $\med$ & $\big[1,\sfrac43\big]$\\
            \typefive & $\med$ & $\big(\sfrac12,\sfrac23\big]$\\
        \hline
    \end{tabular}
        \begin{tabular}{|c|c|c|}
        \hline
            Set & $s_i$ & $s_i + P(C_i)$\\\hline
            \typesix & $\med$ & $\big(\sfrac13,\sfrac49\big]$\\
            \typeseven & $\chp$ & $\big[0,m\big]$\\
        \hline
            \typeeight & $\med$ & $\big(\sfrac49,\sfrac12\big]$\\
            \typenine & $\med \cup \exp$ & $\big(\sfrac23,\sfrac56\big]$\\
            \typeten & $\med \cup \exp$ & $\big(\sfrac56,1\big)$\\
        \hline
    \end{tabular}
    \caption{A total partition of the classes in an instance.}
    \label{tab:preemptive:class_partition}
\end{table}

\subsection{Batch Wrapping}
\label{sec:batch-wrapping}
An algorithmic tool we will use repeatedly is called \emph{Batch Wrapping}, i.e. the wrapping of \emph{wrap sequences}, representing subsets of classes, into \emph{wrap templates}, representing free space on machines, as defined by Deppert and Jansen~\cite{DJ19} and based on ideas of McNaughton~\cite{M59}.
\begin{definition}[Deppert and Jansen~\cite{DJ19}, Wrap Template]
	A \emph{wrap template} is a list $\omega = (\omega_1, \dots, \omega_{\abs{\omega}}) \in (\iv m \times \Q \times \Q)^*$ of triples $\omega_r = (u_r,a_r,b_r) \in \iv m \times \Q \times \Q$ for $1 \leq r \leq \abs{\omega}$ such that $0 \leq a_r < b_r$ and $u_r \leq u_{r+1}$, $\forall 1 \leq r \leq \abs{\omega}$.
\end{definition}
A wrap template thus represents some free space on machines $u_1, \dots, u_{\abs{\omega}}$.
Let $S(\omega) := \sum_{r=1}^{\abs{\omega}} (b_r-a_r)$ denote the total period of time provided by wrap template.
\begin{definition}[Deppert and Jansen~\cite{DJ19}, Wrap Sequence]
	A \emph{wrap sequence} is a sequence
		$$Q = (s_{i_1},j_1^1,\dots,j_{n_1}^1,s_{i_2},j_1^2,\dots,j_{n_2}^2,\quad\dots\quad,s_{i_k},j_1^k,\dots,j_{n_k}^k)$$ 
	where $C'_{i_{\ell}} = \{j_1^{\ell},\dots,j_{n_{\ell}}^{\ell}\}$ with $n_{\ell} = \abs{C'_{i_{\ell}}}$ is a set of jobs and/or job pieces of a class $C_{i_{\ell}}$ for some $i_{\ell} \in [c]$.
\end{definition}
A wrap sequence now represents some subsets of classes in an instance.
Deppert and Jansen~\cite{DJ19} also describe an algorithm \wrap (\autoref{algo:wrap}) that uses wrap templates to schedule wrap sequences.
\begin{algorithm}[h!]
\caption{\wrap$(Q,\omega)$}
\label{algo:wrap}
Schedule all setup times, jobs and job pieces in the order given by $Q$ into tuples $\omega_r$ in the order given by $\omega$, beginning at $a_r$, ending and, if necessary, splitting jobs and job pieces at $b_r$.
Setup times are not split, but instead not scheduled if they reach $b_r$.
For every job or job piece to be scheduled at the beginning of a tuple $\omega_r$, put a setup time of its class directly below $a_r$.
\end{algorithm}
They prove the following properties:
\begin{property}
\label{prop:wrap:space}
Let $Q$ be a wrap sequence containing a largest setup $\smax^{(Q)}$ and $\omega$ be a wrap template with $L(Q) \leq S(\omega)$.
Then, \wrap will place the last job (piece) of $Q$ in a gap $\omega_r$ with $r \leq \abs{\omega}$.
If there was a free time of at least $\smax^{(Q)}$ below each gap but the first, the load gets placed feasibly.
\end{property}
\begin{property}
\label{prop:wrap:rt}
If $L(Q) \leq S(\omega)$, then \wrap$(Q,\omega)$ has a run time of $\Oh(\abs{Q} + \abs{\omega})$.
\end{property}

We give also give a variant of \wrap that is more suited to scheduling larger jobs, which we call \wrapdyn (\autoref{algo:wrapdyn}).
\begin{algorithm}[h!]
\caption{\wrapdyn$(Q,\omega)$}
\label{algo:wrapdyn}
Schedule all setup times, jobs and job pieces in the order given by $Q$ into tuples $\omega_r$ in the order given by $\omega$, beginning at $a_r$, ending and, if necessary, splitting jobs and job pieces at $b_r$.
Setup times are not split, but instead not scheduled if they reach $b_r$.

For every job or job piece to be scheduled at the beginning $a_r$ of a triple $\omega_r$, first schedule a setup time $s_i$ of its class starting at $a_r$, and set the starting time of the job piece to $a_r + s_i$.
\end{algorithm}
\wrapdyn has similar properties to \wrap, which we show in the following.

\begin{property}
\label{prop:wrapdyn:space}
Let $Q$ be a wrap sequence, $\omega$ be a wrap template, and $s_r$ be the first setup time placed on each machine $r \in \abs{\omega}$ by \wrapdyn.
If $L(Q) \leq S(\omega) - \sum_{r=2}^{\abs{\omega}} s_r$ and $b_r - a_r > \max_{r \in \abs{\omega}} s_r$, \wrapdyn will place the last job (piece) of $Q$ in a gap $\omega_r$ with $r \leq \abs{\omega}$.
\end{property}
\begin{proof}
As additional setup times not contained in $Q$ are placed only at the start $a_r$ of any $\omega_r$ for $r > 1$ in the wrap template, which is $s_r$ by definition, the claim follows.
\end{proof}
\begin{property}
\label{prop:wrapdyn:rt}
If $L(Q) \leq S(\omega)$, then \wrapdyn$(Q,\omega)$ has a run time of $\Oh(\abs{Q} + \abs{\omega})$.
\end{property}
To see the main difference between \wrapdyn and \wrap, consider a scenario where there is some interval of size $S$ which we want to use for scheduling a wrap sequence $Q$.
Using \wrap to schedule the jobs, the corresponding entry $\omega_r$ in a wrap template $\omega$ can have size at most $b_r - a_r \leq S - \smax^{(Q)}$ to guarantee $\smax^{(Q)}$ free space below $a_r$, allowing only jobs $j$ with $p_j \leq S - \smax^{(Q)}$ to be placed, with no regard for the actual setup time $s_j$ of $j$.
However, \wrap guarantees that all job pieces are placed above $\smax^{(Q)}$, allowing us to place job pieces of jobs in $Q$ below $\smax^{(Q)}$ without worry of parallelization.
In contrast, using \wrapdyn, for $\omega_r$ only $b_r - a_r \leq S$ must hold, allowing us to place any job $j$ with $s_j + p_j \leq S$ in this interval.
However, $j$ can now start right after $s_j$, \ie below $\smax^{(Q)}$.
We use both \wrap and \wrapdyn in the appropriate places to exploit their respective advantages.
To easier establish feasibility of the schedules produced by \wrapdyn, we additionally show a property regarding parallelization of wrap templates and sequences when using this procedure.
\begin{property}
\label{prop:wrapdyn:para}
If $u_r < u_{r+1}, b_r = b_{r+1}, a_r = a_{r+1}, \forall r \in [1,\abs{\omega}-1]$ and $\max_{\ell \in \iv k, x \in n_{\ell}} s_{i_{\ell}} + j^{\ell}_x \leq b_r - a_r, \forall r \in [\abs{\omega}]$, the schedule produced by \wrapdyn is never parallelized.
\end{property}
\begin{proof}
Let $j^{\ell}_x$ be a job or job piece that starts to be scheduled in $[a_r,b_r]$ by \wrapdyn.
If $j^{\ell}_x$ is scheduled only in $[a_r,b_r]$, the claim follows.
Otherwise, if $j^{\ell}_i$ is interrupted at $b_r$, and then continued to be scheduled in $[a_{r+1},b_{r+1}]$.
As $u_r < u_{r+1}$, $[a_{r+1},b_{r+1}]$ is on a different machine.
Since $s_{i_{\ell}} + j^{\ell}_x \leq b_r - a_r = b_{r+1} - a_{r+1}$ by assumption, and $s_{i_{\ell}}$ is scheduled at $a_{r+1}$ by \wrapdyn, the job piece of $j^{\ell}_i$ scheduled in $[a_{r+1},b_{r+1}]$ is finished processing before the starting time of the job piece of $j^{\ell}_i$ in $[a_r,b_r]$, and the claim thus follows as well.
\end{proof}

\section{Technical Overview}
\label{sec:technical_overview}
For readability purposes, in this section, we informally describe our general procedure to obtain a $(\sfrac{4}{3}+\eps)$-approximation algorithm for \PCmaxpm.
We make use of a modified version of the classical dual approach by Hochbaum and Shmoys~\cite{HS87}, \ie guess a $(1+\eps)$-approximate makespan via binary search in time $\Oh(\log(1/\eps)$, and our algorithm either returns a solution with makespan at most $\sfrac43 T$ for some makespan guess $T$, or allows us to conclude that $T < \OPTpm$ if no feasible solution is returned.

For this, we begin with some structural properties:
We begin by showing that for specific kinds of instances, called \enquote{nice instances} and consisting only of classes of types $\{\typeone, \dots, \typeseven\}$, for any $T$, we can easily (in $\Oh(n)$) compute a $\sfrac43 T$-approximate solution, or conclude that there exists no solution with makespan $T$.
For later arguments, this algorithm places all jobs of $\typeseven$ classes above $\sfrac13 T$.
We then turn our attention to general instances, \ie instances consisting of classes of all types $\{\typeone, \dots, \typeten\}$.
Here, our first goal is to show that for any feasible solution with makespan $T$, there exists a solution with makespan at most $\sfrac{4}{3} T$ that has some specific properties.
We do this by transforming any given solution with makespan $T$ into one with makespan at most $\frac{4}{3} T$ with these properties.
Specifically, we construct a feasible $\sfrac43$ solution by dividing the instance into a nice sub-instance, for which we already know how to compute a $\sfrac43 T$-approximate solution, and \enquote{the rest}.

This rest consists only of classes of $\typeeight, \typenine, \typeten$ and partial classes of $\typeseven$.
For this rest, we transform the given feasible solution with makespan $T$ into one with makespan $\sfrac43 T$ where all classes of $\typenine$ and $\typeten$, and pairs of classes of $\typeeight$, are scheduled together on machines only with one another, or with partial classes of $\typeseven$.
Specifically, jobs of these partial classes of $\typeseven$ are either fully scheduled together with $\typenine$ or $\typeten$ classes in a feasible way, or placed in the interval $[0,\sfrac13 T]$.
This guarantees us that the parts of these $\typeseven$ classes, which may contain job pieces also contained in the nice sub-instance, are never parallelized by the placement all jobs of $\typeseven$ classes above $\sfrac13 T$ of our algorithm for nice (sub-)instances.
Furthermore, the resulting $\frac{4}{3}$-approximate solution is relatively easy to compute.

Finally, we construct a $(\sfrac{4}{3} + \eps)$-approximate solution for any instance by doing the following:
First, we guess the current makespan $T$.
We then construct a solution with makespan at most $\frac{4}{3} T$ and our specified properties in time $\Oh(n^2)$, using the \emph{minimum number of setup times} over all solutions with makespan at most $\frac{4}{3} T$ and the properties as specified above, if a feasible solution exists with makespan $T$.
Thus, if there exists a solution with makespan $T$, we will find a $\frac{4}{3} T$-approximate solution.
If we are not able to compute a feasible $\frac{4}{3} T$-approximate solution, this allows us to conclude that no solution with makespan $T$ exists.
The number of guesses needed to obtain a guess $T$ with $\OPTpm \leq T \leq (1+\eps) \OPTpm$ is $\Oh(\log(1/\eps))$, leading to the total run time of $\Oh(n^2 \log(1/\eps))$.

\section{A \sfrac43-Approximation for Nice Instances}
\label{sec:nice_instances}

In this section we give a \(\sfrac43 T\)-dual approximation for \enquote{nice instances}, \ie instances with some computationally easy to handle properties.
We define nice instances as follows.

\begin{definition}[Nice Instances]
	\label{def:nice_instance}
	For a makespan $T$ we call an instance $I$ \emph{nice} iff \(\abs{\typeeight} \leq 1\) and \(\abs{\typenine} = \abs{\typeten} = 0\).
\end{definition}

\noindent A nice instance \(I\) for a makespan \(T\) has at least an obligatory load of
\begin{align*}
\Lnice(I,T) &\defeq \sum_{i \in \typeone \cup \typetwo} \alpha_i' s_i + \sum_{i \in \typethree} 2 s_i + \sum_{i \in \typefour\cup\typefive\cup\typesix\cup\typeseven\cup\typeeight}s_i + P(J) \\
&\leq \sum_{i \in \iv c} (\alpha_i s_i + P(C_i))
\end{align*} 
where \(P(J)\) is the total processing time of all jobs of the instance, and the last inequality follows from the definition of $\alpha_i$.
For a nice instance \(I\) and a makespan \(T\), we also define $\mnice(I,T)$, which we show to be a lower bound on the minimum machine number necessary to schedule all classes.
Let \(R_3 \subseteq \typethree\), \(R_5 \subseteq \typefive\), and \(R_6 \subseteq \typesix\) such that \(2\floor{\abs{\typethree}/2}+\abs{R_3} = \abs{\typethree}\), \(2\floor{\abs{\typefive}/2}+\abs{R_5} = \abs{\typefive}\), and \(3\floor{\abs{\typesix}/3}+\abs{R_6} = \abs{\typesix}\) for an instance $I$ and a makespan $T$. Moreover, let
    \begin{multline*}
        \mnice(I,T) = \sum_{i \in \typeone \cup \typetwo} \alpha_i' + 3\floor[\Big]{\frac{\abs{\typethree}}2} + \abs{\typefour} + \floor[\Big]{\frac{\abs{\typefive}}2} + \floor[\Big]{\frac{\abs{\typesix}}3} + \\
        \ceil[\Big]{\frac{r_{\op{nice}}(I,T)}T}
    \end{multline*}
    where
    \[
        r_{\op{nice}}(I,T) = \sum_{i \in R_3}\big(2s_i + P(C_i)\big) + \sum_{i \in R_5 \cup R_6 \cup \typeseven \cup \typeeight}\big(s_i + P(C_i)\big).
    \]
If either $I$ and/or $T$ are clear from context, we may omit them and \ie simply write $\Lnice$ or $\mnice$.
We will use both the machine number $\mnice$ as well as the load size $\Lnice$ to argue for the approximate schedulability of a nice instance. The following lemma establishes a relationship between those two values.
\begin{lemma}\label{lemma:lnice_implies_mnice}
Let $I$ be a nice instance and $T$ be a makespan. If $\Lnice \leq mT$, then $\mnice \leq m$.
\end{lemma}
\begin{proof}
    Here we prove the stronger statement that \(\ceil{\frac1T\Lnice} \geq \mnice\). If \(\Lnice \leq mT\), then this gives \(m = \ceil{\frac1TmT} \geq \ceil{\frac1T\Lnice} \geq \mnice\). We may write
    \begin{multline*}
        \Lnice = \sum_{i \in \typeone \cup \typetwo}\Big(\floor[\Big]{\frac{P(C_i)}{T-s_i}}s_i+P(C_i)\Big) + \sum_{i \in \typethree}(2s_i+P(C_i)) + \\
        \sum_{i \in \typefour\cup\typefive\cup\typesix\cup\typeseven\cup\typeeight}(s_i+P(C_i))
    \end{multline*}
    and
    \[
        \mnice = \ceil[\Big]{\sum_{i \in \typeone \cup \typetwo} \floor*{\frac{P(C_i)}{T-s_i}} + 3\floor[\Big]{\frac{\abs{\typethree}}2} + \abs{\typefour} + \floor[\Big]{\frac{\abs{\typefive}}2} + \floor[\Big]{\frac{\abs{\typesix}}3} + \frac{r_{\op{nice}}}T}.
    \]
    Thus, to show \(\ceil{\frac1T\Lnice} \geq \mnice\) it suffices to prove that
    \begin{equation}
        \frac1T\Lnice \geq \sum_{i \in \typeone \cup \typetwo} \floor*{\frac{P(C_i)}{T-s_i}} + 3\floor[\Big]{\frac{\abs{\typethree}}2} + \abs{\typefour} + \floor[\Big]{\frac{\abs{\typefive}}2} + \floor[\Big]{\frac{\abs{\typesix}}3} + \frac{r_{\op{nice}}}T.      
    \label{sufficient-claim}
    \end{equation}
    For all \(i \in \typeone\cup\typetwo\) we find
    \begin{align*}
        \frac1T\Big(\floor[\Big]{\frac{P(C_i)}{T-s_i}}s_i+P(C_i)\Big)
        &= \frac1T\Big(\floor[\Big]{\frac{P(C_i)}{T-s_i}}s_i+\frac{P(C_i)}{T-s_i}(T-s_i)\Big)\\
        &\geq \frac1T\Big(\floor[\Big]{\frac{P(C_i)}{T-s_i}}s_i+\floor[\Big]{\frac{P(C_i)}{T-s_i}}(T-s_i)\Big)\\
        &= \floor[\Big]{\frac{P(C_i)}{T-s_i}}
    \end{align*}
    and
    \begin{align*}
        \frac1T\sum_{i \in \typethree \setminus R_3}(\underbrace{s_i}_{>\,\frac13T}+\underbrace{s_i+P(C_i)}_{>\,\frac43T})
        &\geq \frac53\abs{\typethree \setminus R_3}
        = \frac53(\abs{\typethree}-\abs{R_3})\\
        &= \frac53\cdot 2\floor[\Big]{\frac{\abs{\typethree}}2}
        \geq 3\floor[\Big]{\frac{\abs{\typethree}}2}
    \end{align*}
    and
    \[
        \frac1T\sum_{i \in \typefour}(\underbrace{s_i+P(C_i)}_{>\,T})
        \geq \abs{\typefour}
    \]
    and
    \begin{align*}
        \frac1T\sum_{i \in \typefive \setminus R_5}(\underbrace{s_i+P(C_i)}_{>\,\frac12T})
        &\geq \frac12\abs{\typefive \setminus R_5}
        = \frac12(\abs{\typefive}-\abs{R_5})\\
        &= \frac12 \cdot 2\floor[\Big]{\frac{\abs{\typefive}}2}
        = \floor[\Big]{\frac{\abs{\typefive}}2}
    \end{align*}
    and
    \begin{align*}
        \frac1T\sum_{i \in \typesix \setminus R_6}(\underbrace{s_i+P(C_i)}_{>\,\frac13T})
        &\geq \frac13\abs{\typesix \setminus R_6}
        = \frac13(\abs{\typesix}-\abs{R_6})\\
        &= \frac13 \cdot 3\floor[\Big]{\frac{\abs{\typesix}}3}
        = \floor[\Big]{\frac{\abs{\typesix}}3}.
    \end{align*}
    By comparing the sides by type, this shows \autoref{sufficient-claim}, which proves the claim.
\end{proof}
We are now ready to describe the $\sfrac43$-approximation algorithm for a nice instance $I$.
Note that the algorithm places all jobs of classes in \typeseven only above $\sfrac13T$, which will be important for later reasoning.
\begin{theorem}\label{preemptive:simple}
    Let $I$ be a nice instance with optimal objective value $\OPTpm$, and $T$ be a makespan.
    Then the following properties hold.
	\begin{enumerate}[(i)]
		\item If $m < m_{\op{nice}}$, it is true that $T < \OPTpm$.
		\label{preemptive:simple:decision:T_check_false}
		\item Otherwise a feasible schedule with makespan at most $\tfrac43T$ can be computed in time $\Oh(n)$.
		\label{preemptive:simple:decision:T_check_true}
	\end{enumerate}
\end{theorem}
\begin{proof}
(\ref{preemptive:simple:decision:T_check_false}).
We show that $T \geq \OPTpm$ implies $mT \geq \Lnice$, which in turn implies $m \geq \mnice$ (\autoref{lemma:lnice_implies_mnice}).
Let $T \geq \OPTpm$.
Then there exists a feasible schedule $\sigma$ with makespan $T$.
Let $L(\sigma) = \sum_{i = 1}^c (\lambda^{\sigma}_i s_i + P(C_i))$.
Then
$$mT \geq L(\sigma) \geq \sum_{i = 1}^c \alpha_i s_i + P(J) \geq \Lnice \enspace ,$$
where the second inequality follows from \autoref{lemma:lambda_geq_alpha}.

(\ref{preemptive:simple:decision:T_check_true}).
Here, $m \geq \mnice$.
Consider the following procedure:
\label{algo:nice}
\begin{enumerate}
    \item Schedule all $i \in \typeone \cup \typetwo$ on  $\alpha_i'$ machines each, by scheduling $s_i$ setup times starting at $0$ on $\alpha_i'$ machines, and the jobs in arbitrary order one after another on the $\alpha_i'$ machines between $s_i$ and $1 + (P(C_i) \mod (T-s_i)) / \alpha_i'$, for each $i$, preempting each job at $1 + (P(C_i) \mod (T-s_i)) / \alpha_i'$ and restarting it at $s_i$ on the next machine if necessary. 
    \item Schedule pairs $i,i' \in \typethree$ together on $3$ machines: 
    Schedule $s_i$ and $P(C_i)-(\sfrac{2}{3}-s_i)$ processing time of $i$ on the first machine, $s_{i}$ and $\sfrac{2}{3}-s_{i}$ processing time of $C_{i}$, as well as $s_{i'}$ and $\sfrac{2}{3}-s_{i'}$ processing time of $i'$ on the second machine, and $s_{i'}$ and $P(C_{i'})-(\sfrac{2}{3}-s_{i'})$ processing time of $i'$ on the third machine.
    \item Schedule each $i \in \typefour$ on a single machine.
    \item Schedule all classes in $\typefive$ as pairs $i,i'$ together on a single machine.
    \item Schedule all classes in $\typesix$ as triples  $i,i',i''$ together on a single machine.
\end{enumerate}

After these steps, there is at most one class $i_3 \in \typethree$, one class $i_5 \in \typefive$, two classes $i_6, i'_6 \in \typesix$, at most one class $\ieight \in \typeeight$, and all classes in $\typeseven$ unscheduled in a nice instance.
We first deal with all but the $\typeseven$ classes.
If $i_3$ exists, always schedule $s_3$ and $P(C_3)-(\sfrac{2}{3}-s_3)$ processing time of $i_3$ on a single machine; this leaves exactly $P'(C_3) = \sfrac23 - s_3$ processing time of $i_3$ unscheduled.
Let $P'(C_i) = P(C_i)$ for all $i \in \{i_5, i_6, i'_6, \ieight\}$.
Then, $s_i + P'(C_i) + s_{i'} + P'(C_{i'}) \leq \sfrac43$, for all $i,i' \in \{i_3, i_5, i_6, i'_6, \ieight\}$.
Thus, for any combination of $0$ to $4$ classes of $\{i_3, i_5, i_6, i'_6, \ieight\}$, we can schedule the remaining processing time of these classes as given by $P'$ on at most $2$ machines.
If all $5$ classes in  $\{i_3, i_5, i_6, i'_6, \ieight\}$ are yet unscheduled, we schedule $1 < s_3 + P'(C_3) + s_6 + P'(C_6) \leq \sfrac43$ on one machine; the remaining processing time of the remaining classes $\{i_5, i'_6, \ieight\}$ can then be scheduled again on at most $2$ machines as before.

We do exactly this, scheduled such that one machine $q_s$ starts processing two classes of $\{i_5, i_6, \allowbreak i'_6, \ieight\}$ at $0$, finishing at some time $a \leq \sfrac43$, and the other machine $q_e$ finishes processing two classes of $\{i_5, i_6, i'_6, \ieight\}$ at $\sfrac43$, starting at some time $b \geq 0$.
Let $\Mrem$ be the set of all machines with nothing yet scheduled on them.
If $L(q_s) + L(q_e) \leq \sfrac43$, we instead schedule all classes of $\{i_5, i_6, i'_6, \ieight\}$ together on $q_s$ starting at $0$ and finishing at $a \leq \sfrac43$, leaving $q_e$ empty and setting $b = \sfrac43$.
To now schedule all classes $i \in \typeseven$, we create a Wrap Template $\omega = (\omega_1, \dots, \omega_{\abs{\Mrem}+2}$, with $\omega_1 = (q_s, \max(a,\sfrac13 T), \sfrac43 T)$, $\omega_{\abs{\Mrem}+2} = (q_e, \sfrac13 T, b)$, and $\omega_r = (q_{r-1}, \sfrac13, \sfrac43)$ for all $r = {2, \dots, \abs{\Mrem}+1}$, with $q_r$ being the $r$-th machine in $\Mrem$, as well as a Wrap Sequence $Q$ with an entry $s_i,j_1^i,\dots,j_{n_i}^i$ for each $i \in \typeseven$, $C_i = \{j_1^i, \dots, j_{n_i}^i\}$ in arbitrary order.
Finally, we execute \wrap($Q,\omega$).

There are enough machines to execute all steps without fail: As we use at most $\alpha_i$ setup times while scheduling each class $i \in \typeone \cup \dots \cup \typesix \cup \typeeight$ to schedule these classes on some $m'$ machines, distributing $> m' T$ load in the process, if at some point, there was no machine to schedule $i$, $m < \mnice$ by definition of $\mnice$, a contradiction.
All $m'$ machines finish for before $\sfrac43 T$: For $i \in \typeone$, $1 + (P(C_i) \mod (T-s_i)) / \alpha_i' \leq \sfrac43$, as $s_i \geq \sfrac23 T$, and for $i \in \typetwo$, $1 + (P(C_i) \mod (T-s_i)) / \alpha_i' \leq \sfrac43$ as well, as  $s_i \geq \sfrac13 T$ and $\alpha_i' \geq 2$.
For $i \in \typethree \cup \dots \cup \typesix \cup \typeeight$, each machine finishes before $\sfrac43 T$ by definition of the procedure. 
None of the jobs scheduled in this way are parallelized, as all jobs are either never preempted (for classes in \typefour, \typefive, \typesix and \typeeight), or preempted only at some time $> T$, and then continued at $s_i$ until they finish (for classes in \typeone, \typetwo and \typethree), making parallelization impossible as $p_j \leq T - s_i$ for all jobs $j$ in the instance.
All these steps can be computed in total time $\Oh(n)$.

All classes in $\typeseven$ are included in $Q$ and scheduled via \wrap($Q,\omega$).
As $S(\omega) \geq mT - \sum_{i \in \{\typeone, \dots, \typesix, \typeeight\}} s_i + P(C_i) \geq \mnice T - \sum_{i \in \{\typeone, \dots, \typesix, \typeeight\}} s_i + P(C_i) \geq \sum_{i \in \typeseven} s_i + P(C_i) = L(Q)$, \autoref{prop:wrap:rt} is fulfilled.
As $\smax^{(Q)} \leq \sfrac13$, and there is free time of $\sfrac13$ for each $\omega_r$ with $r > 1$ by definition of $\omega$, \autoref{prop:wrap:space} is fulfilled as well.
The minimal continuous, non-overlapping free space in $\omega$ is $\min(S(\omega),1)$; the largest processing time $p_j$ of any job $j \in Q$ is at most $\min(L(Q),1-s_j)$; therefore, the partial schedule created by \wrap($Q,\omega$) is never parallelized.

\begin{figure}[h]
    \centering
    \begin{adjustbox}{width=\linewidth, keepaspectratio}
    \tikz[xscale=0.8,yscale=2.5]{
        \tikzmath{\m = 19;}
        \hlines \m {
            {4/3}/$\frac43$//,
            1/$1$/solid/,
            {2/3}/$\frac23$//,
            {1/3}/$\frac13$//,
            0/$0$/solid/
        };
        \schedule up {
            {
                .51//fill=gray/,
                .78//fill=lightgray/
            }/1/,
            {
                .61//fill=gray/,
                .55//fill=lightgray/
            }/1/,
            {
                .37//fill=gray/,
                .22//fill=lightgray/,
                .39//fill=gray/,
                .27//fill=lightgray/
            }/1/,
            {
                .42//fill=gray/,
                .24//fill=lightgray/,
                .45//fill=gray/,
                .10//fill=lightgray/
            }/1/,
            {
                .34//fill=gray/,
                .04//fill=lightgray/,
                .38//fill=gray/,
                .06//fill=lightgray/,
                .35//fill=gray/,
                .09//fill=lightgray/
            }/1/,
            {
                .35//fill=gray/,
                .08//fill=lightgray/,
                .34//fill=gray/,
                .04//fill=lightgray/,
                .38//fill=gray/,
                .05//fill=lightgray/
            }/1/,
            {
                .88//fill=gray/,
                .32//fill=lightgray/
            }/1/,
            {
                .88//fill=gray/,
                .32//fill=lightgray/
            }/1/,
            {
                .58//fill=gray/,
                .68//fill=lightgray/
            }/1/,
            {
                .58//fill=gray/,
                .68//fill=lightgray/
            }/1/,
            {
                .58//fill=gray/,
                .68//fill=lightgray/
            }/1/,
            {
                .36/$s_i$/fill=gray/,
                .70//fill=lightgray/
            }/1/,
            {
                .36/$s_i$/fill=gray/,
                {2/3-.36}//fill=lightgray/,
                .56/$s_j$/fill=gray/,
                {2/3-.56}//fill=lightgray/
            }/1/,
            {
                .56/$s_j$/fill=gray/,
                .70//fill=lightgray/
            }/1/,
            {
                .44/$s_k$/fill=gray/,
                .70//fill=lightgray/
            }/1/,
            {
                .44/$s_k$/fill=gray/,
                {2/3-.44}//fill=lightgray/,
                .22//{preaction={fill,gray},pattern=north east lines}/,
                {2/3-.22}//{preaction={fill,lightgray},pattern=north east lines}/
            }/1/,
            {
                {1/3-.22}//opacity=0/,
                .22//{preaction={fill,gray},pattern=north east lines}/,
                .83//{preaction={fill,lightgray},pattern=north east lines}/,
                .13//{preaction={fill,gray},pattern=north east lines}/,
                {.04}//{preaction={fill,lightgray},pattern=north east lines}/
            }/1/,
            {
                {1/3-.13}//opacity=0/,
                .13//{preaction={fill,gray},pattern=north east lines}/,
                .27//{preaction={fill,lightgray},pattern=north east lines}/,
                .13//{preaction={fill,gray},pattern=north east lines}/,
                .11//{preaction={fill,lightgray},pattern=north east lines}/,
                .18//{preaction={fill,gray},pattern=north east lines}/,
                .15//{preaction={fill,lightgray},pattern=north east lines}/
            }/1/
        };
        \draw [|-|] (0,-.1) -- node {\typefour} (2,-.1);
        \draw [|-|] (2,-.1) -- node {\typefive} (4,-.1);
        \draw [|-|] (4,-.1) -- node {\typesix} (6,-.1);
        \draw [|-|] (6,-.1) -- node {\typeone} (8,-.1);
        \draw [|-|] (8,-.1) -- node {\typetwo} (11,-.1);
        \draw [|-|] (11,-.1) -- node {\typethree} (16,-.1);
        \draw [|-|] (15,-.15) -- node {\typeseven} (18,-.15);
    }
    \end{adjustbox}
    \caption{A $\sfrac43$-approximate solution of a nice instance.}
    \label{fig:solving-friendly-instances}
\end{figure}
\end{proof}

\section{Structural Insights on General Instances}
\label{sec:structure}
In the previous section, we have seen how to efficiently compute $\sfrac43$-approximate solutions for so-called \enquote{nice instances}, \ie instances with arbitrary many classes only in types $\typeone \cup \dots \cup \typeseven$, and at most one class in $\typeeight$.
We turn our attention now to general instances.
In this section, we begin by showing some structural properties of such instances.
The goal is to prove that, for every instance, if there exists a feasible solution for some makespan guess $T$, there also exists a $\sfrac43 T$-approximate solution with some nice properties.
Specifically, we aim to show that there every instance can be split into a nice instance and \enquote{the rest}.
For the nice instance, the $\sfrac43 T$-approximate solution is easy to compute; the rest will be shown to have some easy to compute properties as well.
We do this by taking any optimal solution and successively transforming it into a $\sfrac43 T$-approximate solution with the desired properties.
In the algorithmic section, the hard part then becomes to compute such a solution without prior knowledge of the optimal, $T$-feasible solution.

For an instance $I$ of \PCmaxpm, let $\solopt$ be an optimal solution with makespan $T$.
For readability purposes, we normalize the instance by dividing all relevant input parameters by $T$.
At various points, we will combine even multiples of classes of \typeeight on even multiples of machines containing such classes.
If, at such a step, the number of \typeeight classes to be combined is odd, we add the remaining \typeeight class, all setup times and job pieces scheduled on the same machine $q$ as this class, as well as the machine $q$ itself to a sub-instance $\Ieightrem$, which we will deal with afterwards.

In $\solopt$, there is a subset of machines where only classes of types $\typeone \cup \dots \cup \typeseven$ scheduled on them.
We call this set of machines $\Mnice$, and the subset of setup times and job pieces scheduled on these machines $\Snice$.
Let $\Inice$ be the sub-instance induced by $\Mnice$ and all setup times and job pieces of jobs in $\Snice$.
By definition of $\solopt$, it holds for $\Inice$ that $L(\Snice) \leq \abs{\Mnice}$.
Note that $\Inice$ is not necessarily a nice instance yet, as even tough it only consists of setup times and job pieces of nice classes from the original instance, not necessarily all job pieces of each nice class are included in $\Inice$.
This may change the types of classes in the sub-instance.
Specifically, each class $i \in \typeone \cup \dots \cup \typesix$ hitherto partially included in $\Inice$, we aim to include in $\Inice$ completely: This preserves the type of $i$ for fixed makespan $\solopt$.
For classes $i \in \typeseven$, we may include only some job pieces of $i$ in $\Inice$ without violating the definition of a nice instance: Each subset of a class $i \in \typeseven$ is itself in $\typeseven$.

\subsection{Non-Nice Classes with Multiple Setup Times}

Now we begin to show how to also deal with machines containing non-nice types of classes.
We start with those that use multiple setup times in $\solopt$.
Let $M_{\lambda \geq 2}$ be the subset of machines in $\solopt$ with classes $i \in \typeeight \cup \typenine \cup \typeten$ scheduled on them and $\lambda^{\solopt}_i \geq 2$, \ie all machines with classes of types \typeeight, \typenine or \typeten that use more than one setup in $\solopt$.
We first remove all setup times and job pieces from $M_{\lambda \geq 2}$, and then schedule all such classes of \typeeight as pairs on one machine, and each such class of \typenine and \typeten alone on one machine of $M_{\lambda \geq 2}$, placing all these classes of \typeeight, \typenine and \typeten. %
Let $M'_{\lambda \geq 2} \subseteq M_{\lambda \geq 2}$ be the set of machines used in the process of rescheduling above.
Then the total sum $\Lrest$ of all setup times and job pieces $\Srest$ scheduled on $M_{\lambda \geq 2}$ in $\solopt$, but not scheduled on $M'_{\lambda \geq 2}$ is at most $L(M_{\lambda \geq 2}) - L(M'_{\lambda \geq 2}) - \sfrac13 \abs{M'_{\lambda \geq 2}}$, as for each class $i \in \typeeight \cup \typenine \cup \typeten$ scheduled on $M'_{\lambda \geq 2}$, there is now exactly one setup time scheduled on a machine in $M'_{\lambda \geq 2}$, but there were at least $\lambda^{\solopt}_i \geq 2$ setup times scheduled on $M_{\lambda \geq 2}$ scheduled in $\solopt$; by definition of the classes $i$, each such setup time is $> \sfrac13$.

By definition of $\solopt$, $L(M_{\lambda \geq 2}) \leq \abs{M_{\lambda \geq 2}}$.
As it now holds for the load of each machine $q \in M'_{\lambda \geq 2}$ that $L(q) \geq \sfrac23$ by definition of the types scheduled on them, we can conclude that $\Lrest = L(M_{\lambda \geq 2}) - L(M'_{\lambda \geq 2}) - \sfrac13 \abs{M'_{\lambda \geq 2}} \leq \abs{M_{\lambda \geq 2}} - \sfrac23 \abs{M'_{\lambda \geq 2}} - \sfrac13 \abs{M'_{\lambda \geq 2}} = \abs{M_{\lambda \geq 2}} - \abs{M'_{\lambda \geq 2}}$.
All setup times and job pieces in $\Lrest$ are from classes $i \in \typeone \cup \dots \cup \typeseven$ by definition.
We set $\Snice = \Snice \cup \Srest$ and $\Mnice = \Mnice \cup (M_{\lambda \geq 2} \setminus M'_{\lambda \geq 2})$, preserving $L(\Snice) \leq \abs{\Mnice}$.

\subsection{Constructing a Nice Sub-Instance}
After dealing with all non-nice types of classes that use multiple setup times in $\solopt$, we are only now left to deal with the non-nice types that use exactly one setup time in $\solopt$.
Let $M_{\lambda = 1}$ be the set of machines with classes $i \in \typeeight \cup \typenine \cup \typeten$ and $\lambda^{\solopt}_i = 1$.

Note that $M = \Mnice \cupdot M_{\lambda = 1} \cupdot M'_{\lambda \geq 2}$, and that for all $q \in M_{\lambda = 1}$ that have a class $i \in \typenine \cup \typeten$ scheduled on them in $\solopt$, all remaining setup times and job pieces in $\solopt$ belong to classes in $\typeseven$: The setup times of all other types of classes are already to large to be scheduled together with $i$ on $q$.
Let $\Mnine \subseteq M_{\lambda = 1}$ be the set of machines with a class of \typenine, $\Mten \subseteq M_{\lambda = 1}$ the set of machines with a class of \typeten scheduled on them in $\solopt$.
For all machines $q \in M_{\lambda = 1}$ that have at least class $i \in \typeeight$ scheduled on them in $\solopt$, there is either another class $i' \in \typeeight$ scheduled on $q$, or not.
Let $\Mdual \subseteq M_{\lambda = 1}$ be the set of machines for which the former, $\Msingle \subseteq M_{\lambda = 1}$ the set of machines for which the latter is true.
All remaining setup times and job pieces in $\solopt$ scheduled on machines in $\Mdual$ belong to classes in $\typeseven$ by the same argument as above.
All remaining setup times and job pieces in $\solopt$ scheduled on machines in $\Msingle$ belong to classes in $\typeone \cup \dots \cup \typeseven$ by definition of the machine set.
Note that $M_{\lambda = 1} = \Msingle \cupdot \Mdual \cupdot \Mnine \cupdot \Mten$.

We first want to finish making the sub-instance induced by $\Snice$ and $\Mnice$ an actual nice instance.
This may not be the case yet: For some classes $i \in \typeone \cup \dots \cup \typesix$, there may be some job pieces not yet included in $\Snice$, meaning that in the sub-instance, these $i$ are not complete.
The only job pieces of such classes not included in $\Snice$ must be scheduled on machines $q \in \Msingle$, as argued above.
We now include these job pieces, and all other classes that are scheduled over multiple machines of $\Msingle$ in $\solopt$, in $\Snice$.
Let $\Msingle' \subseteq \Msingle$ be the set of machines where at least one setup time and a job piece of a class $i \in \typeone \cup \dots \cup \typesix$ is scheduled in $\solopt$, but not the whole class, \ie $\Msingle' = \{q \in \Msingle \vert \exists i \in \typeone \cup \dots \cup \typesix, j \in C_i : \exists p_j: \solopt(j) = q, P(C_i) > p_j > 0 \}$.
Note that the rest of class $i$ may be scheduled on machines $q' \in \Msingle'$ or included in $\Snice$.

Now, we do the following modifications:
While there exists a class $i \in \typeone \cup \dots \cup \typesix$ such that two job pieces $p_j,p'_j \in C_i$ are scheduled on two different machines $q,q' \in \Msingle'$, schedule both classes $\ieight,\ieight'$ previously scheduled on $q,q'$ together on $q$, and add all job pieces of class $i$ and all setup times and job pieces not belonging to class $i$ that have been previously scheduled on $q$ and $q'$, as well as one setup time $s_i$ to $\Snice$, and add $q'$ to $\Mnice$.
The machine $q$ is added to $\Mdual$.
The total load of the setup times and job pieces added to $\Snice$ in every step is at most $\sfrac79$, as the total load of all setup times in job pieces on machines $q$ and $q'$ was at most $2$ in $\solopt$, since we did both not include classes $\ieight,\ieight'$ of load at least $\sfrac49$ each, as well as discard one of the at least two setup times $s_i \geq \sfrac13$ that were previously scheduled on $q$ and $q'$ as well.
As $q'$ is added to $\Mnice$, $L(\Snice) \leq \abs{\Mnice}$ is preserved.

Now, for each class $i \in \typeone \cup \dots \cup \typesix$ with more than one setup time $s_i$ in $\solopt$, there is at most one machine $q \in \Msingle'$ with job pieces of $i$ scheduled on it in $\solopt$.
While there exists two different such classes $i,i'$, with job pieces on machines $q,q' \in \Msingle'$, we again schedule both classes $\ieight,\ieight'$ previously scheduled on $q,q'$ together on $q$ and add $q$ to $\Mdual$.
We add all job pieces pieces of classes $i$ and $i'$, but none of their setup times, and all setup times and job pieces not belonging to classes $i$ or $i'$ that have been previously scheduled on $q$ and $q'$ to $\Snice$, as well as the machine $q'$ to $\Mnice$.
The total load added to $\Snice$ is at most $2-2(\sfrac49) - 2(\sfrac13) = \sfrac29$ by the same arguments as above, and by adding $q'$ to $\Mnice$, $L(\Snice) \leq \abs{\Mnice}$ is preserved.
Afterwards, if $\abs{\Msingle'} \neq 0$, $\abs{\Msingle'} = 1$ by the pairing above,
and we remove $q \in \Msingle'$ from this set, add the class $\ieight$ of \typeeight on $q$ to $\Ieightrem$, and add all setup times and job pieces but $\ieight$ scheduled on $q$ in $\solopt$ to $\Snice$ and $q$ to $\Mnice$.

Now, $\Inice$ actually constitutes a nice instance: All classes of types $\typeone \cup \dots \cup \typesix$ that are included in $\Snice$ are included fully.
It is only left to argue that $\Lnice(\Inice) \leq \abs{\Mnice}$, which implies that we can compute a $\sfrac43$-approximation for this sub-instance in time $\Oh(n)$ by way of \autoref{lemma:lnice_implies_mnice} and \autoref{preemptive:simple},~(\ref{preemptive:simple:decision:T_check_true}).
Since we already know that $L(\Snice) \leq \abs{\Mnice}$, it is enough to show $\Lnice(\Inice) \leq L(\Snice)$. Let $\lambda'_i$ be the number of setup times of class $i$ in $\Snice$, and $J$ the set of all job pieces in $\Snice$.
$\Lnice(\Inice) = P(J) + \sum_{i \in \typeone \cup \typetwo \cup \typefour} \alpha_i' s_i + \sum_{i \in \typethree \cup \typefive \cup \typesix} \alpha_i s_i + \sum_{i \in \typeseven} s_i \leq P(J) + \sum_{i \in \typeone \cup \dots \cup \typeseven} \lambda'_i = L(\Snice)$ is true if,

$$\alpha_i' \leq \lambda'_i, \forall i \in \typeone \cup \typetwo \cup \typefour \enspace , \enspace
\qquad \alpha_i \leq \lambda'_i, \forall i \in \typethree \cup \typefive \cup \typesix \enspace ,$$

$$\text{ and} \enspace 1 \leq \lambda'_i, \forall i \in \typeseven \enspace .$$

For all $i \in \typeone$, there actually is no job part on any machine $q \in \Msingle'$ in $\solopt$, as the setup of any such class and a full class of $\typeeight$ do not fit together on a single machine.
Thus, no setup of $i$ is removed, and $\lambda'_i = \lambda^{\solopt}_i \geq \alpha'_i$.
For all $i$, let $P'(C_i)$ be the load originally not scheduled on machines in $\Msingle'$ in $\solopt$, and $P''(C_i)=P(C_i)-P'(C_i)$ be the load that is.
Then, for all $i \in \typetwo$, the total number of setup times of $i$ in $\solopt$ is at least $\ceil*{P'(C_i)/(1-s_i)} + \floor*{P''(C_i)/(\sfrac{10}{9}-2s_i)} > \floor*{P'(C_i)/(1-s_i)} + \floor{P''(C_i)/(1-s_i)} = \alpha'_i$, as there is at most $(\sfrac{10}{9}-2s_i)$ of $P''(C_i)$ on two machines in $\Msingle'$, and there is a setup time $s_i$ added to $\Snice$ for every such two machines. The last inequality follows from the fact that $(1-s_i) > (\sfrac{10}{9}-2s_i)$ for $s_i \geq \sfrac13$, which is the case for $i \in \typetwo$.
For $i \in \typethree$, $\alpha_i = 2$ by definition.
If $P'(C_i) \geq (1-s_i)$, $\lambda'_i \geq 2$ follows immediately.
Otherwise, $P''(C_i) > \sfrac13$ by definition of $\typethree$, and thus is scheduled over at least $2$ machines in $\Msingle$ if $P'(C_i) > 0$, or $P''(C_i) > 1$ if $P'(C_i) =  0$ and thus scheduled over at least $4$ machines in $\Msingle$, as at most $1 - \sfrac49 - \sfrac13 = \sfrac29$ processing time of $i$ can be scheduled on a machine in $\Msingle$.
In any case, $\lambda'_i \geq 2$ in $\solopt$.
For all $i \in \typefour$, $\lambda'_i \geq \alpha'_i$, as $\alpha_i' = 1$ and $\lambda'_i \geq 1$ by definition.
The same holds for all $i \in \typefive \cup \typesix$ and $\lambda'_i \geq \alpha_i$, as $\alpha_i=1$ for such $i$, as well as for all $i \in \typeseven$.
Thus, $\Inice$ is not only a nice instance, but we can calculate a $\sfrac43$-approximation in time $\Oh(n)$ for it.

\subsection{Combining Classes of Type Eight and Nine}

Now, for each remaining machine in the set $M_{\lambda=1}$, we not only know that there is either exactly one class $i \in \typeeight \cup \typenine \cup \typeten$ or exactly two classes $i \in \typeeight$ fully scheduled on it in $\solopt$, but also that for each machine with one $i \in \typenine \cup \typeten$ or two $i,i' \in \typeeight$ on it, all other job pieces scheduled on this machine belong to some classes of $\typeseven$;
and for each machine with one $i \in \typeeight$ on it, there is at most one class $i' \in \typefive \cup \typesix$ fully scheduled on it (as no other nice class fully fits on such a machine in $\solopt$), and all other job pieces scheduled on this machine belong to some classes of $\typeseven$.
We proceed by combining classes of \typeeight and \typenine efficiently.
While $\Msingle \neq \emptyset$ and $\Mnine \neq \emptyset$, let $q \in \Msingle$, $q' \in \Mnine$ and $\ieight$ the $\typeeight$-class on $q$, $\inine$ the $\typenine$-class on $q'$.
Schedule both $\ieight$ and $\inine$ on $q$.
The new load of $q$ then is always at least $\sfrac49 + \sfrac23 = 10/9 > 1$, and at most $\sfrac12 + \sfrac56 = \sfrac43$.
Thus, the load of all setup times and job pieces previously scheduled on $q$ and $q'$ without $\ieight$ and $\inine$ is at most $2 - 10/9 < 1$.
As all those setup times and job pieces amount to either full classes of $\typefive \cup \typesix$ and/or are job pieces of some classes of $\typeseven$, adding them to $\Snice$ and $q'$ to $\Mnice$ preserves both the nice property of $\Inice$, as well as $\Lnice(\Inice) \leq \abs{\Mnice}$.
If after this process, $\Mnine = \emptyset$, we are done.
Otherwise, $\Msingle = \emptyset$.
Now, while $\Mdual \neq \emptyset$ and $\abs{\Mnine} \geq 2$, let $q \in \Mdual$, $q',q'' \in \Mnine$ and $\ieight,\ieight'$ the $\typeeight$-class on $q$, $\inine,\inine'$ the $\typenine$-classes on $q',q''$, respectively.
Similarly, schedule both $\ieight$ and $\inine$ on $q$, $\ieight'$ and $\inine'$ on $q'$.
Then the new load of both $q$ and $q'$ is always at least $\sfrac49 + \sfrac23 = 10/9 > 1$, and at most $\sfrac12 + \sfrac56 = \sfrac43$.
With the same argument as above, we add the remaining setup times and job pieces previously also scheduled on $q,q',q''$ to $\Snice$, and $q''$ to $\Mnice$, preserving the aforementioned properties.
If, after this, $\Mnine \neq \emptyset$ and $\Mdual \neq \emptyset$, then $\abs{\Mnine} = 1$.
For the remaining \typenine-class $\inine$ on a machine $q \in \Mnine$, we schedule it together on $q$ with a \typeeight-class $\ieight$ taken from a machine $q' \in \Mdual$.
We leave all other setup times and job pieces on $q'$, but schedule all other setup times and job pieces, all of which belong to classes of $\typeseven$, previously scheduled on $q$ on $q'$, and add $q'$ to $\Msingle$.
The resulting schedule might not be feasible anymore, but we will restore feasibility later while dealing with all machines in $\Msingle$.
Note that now, $L(q') \leq 1 - \sfrac49 + \sfrac13 = \sfrac89$ by definition of the removed $\ieight$, and the total load of setup times and jobs removed from $q$ and put on $q'$.

Now, either $\Mnine = \emptyset$ or $\Msingle = \Mdual = \emptyset$, \ie we have either dealt with all $\typenine$ or all $\typeeight$ classes.
Since \typeten is the only other non-nice type to be distributed, we make a case distinction: We show how to distribute classes of \typeten together only with classes of \typeeight, or together only with classes of \typenine.
As either case will also applicable if neither classes of \typeeight or \typenine are remaining, this then shows how to distribute all non-nice types.

\subsection{Classes of Type Eight Remain}
\label{sec:eight_rem}
We start by constructing a bipartite graph, with a vertex for each class of \typeeight on $\Msingle \cup \Mdual$ and \typeten in $\Mten$, and an edge between two vertices iff one vertex represents a class of \typeeight, the other a class of \typeten, and the combined load both classes is lesser or equal than $\sfrac43$.
We compute a maximal matching for this graph, iteratively scheduling all classes of $\ieight \in \typeeight$ and $\iten \in \typeten$ corresponding to the vertices of an edge included in the maximal matching together on the machines $q' \in \Mten$ where $\iten$ was scheduled previously, and remove $q'$ from $\Mten$.
If $\ieight$ was previously scheduled on a machine $q \in \Msingle$, all setup times and job pieces of previously scheduled on $q$ and $q'$, except for $\ieight$ and $\iten$, are added to $\Snice$, and $q$ is removed from $\Msingle$, and added to $\Mnice$.
As all of these setup times and job pieces are either parts of \typeseven classes, or include a full \typefive or \typesix class, and their total load is at most $2-(\sfrac56 + \sfrac49) = \sfrac{13}{18} < 1$ by the definition $\ieight$ and $\iten$, this preserves $\Lnice(\Inice) \leq \abs{\Mnice}$.

If $\ieight$ was instead previously scheduled on a machine $q \in \Mdual$, we schedule all setup times and job pieces of previously scheduled on $q$ and $q'$, except for $\ieight$ and $\iten$, on $q$, removing $q$ from $\Mdual$ and adding it to $\Msingle$.
The resulting schedule might not be feasible anymore, but we will restore feasibility later while dealing with all other machines in $\Msingle$.
Again, by definition of $\ieight$ and $\iten$, $L(q) \leq 2-(\sfrac56 + \sfrac49) = \sfrac{13}{18}$.
By definition of the graph and the minimum load of classes $\ieight$ and $\iten$, respectively, machine $q'$ now has load at least $1$ and at most $\sfrac43$.
It then follows for all combinations of remaining classes of \typeeight and \typeten after this process, that their combined load is more than $\sfrac43$: Otherwise, the matching would not have been maximal.

Now consider the remaining machines with classes of \typeeight and \typeten with their scheduling not modified by the process above.
We start by adding all setup times and job pieces scheduled on machines $q \in \Mdual \cup \Mten$, the \typeeight, \typeten and (partial) \typeseven classes, to a set $\Seight$, which will store all setup times in job pieces not included in the nice sub-instance, and set $\Meight = \Mdual \cup \Mten$, which represents all machines not yet used at another time in the modified schedule.
Later, we will show that we can use a combination of direct placement and Batch-Wrapping to place all these setup times and job pieces in $\Seight$ on the machines of $\Meight$ in a feasible way.
We proceed by looking at $\Msingle$.
We aim to combine two \typeeight classes on machines in $\Msingle$, with the addition of some job pieces of \typeseven classes, on one of their machines, such that the other machine and the remaining job pieces can be added to $\Inice$ while preserving computability of the nice $\sfrac43$-approximation.
Let $\Srest$ be the set of setup times and job pieces on $q$ and $q'$ not belonging to $\ieight$ or $\ieight'$, \ie some (partial) \typeseven classes, for some $q,q' \in \Msingle$ and $\ieight$ scheduled on $q$, $\ieight'$ scheduled on $q'$ in $\solopt$.
We first consider some \enquote{easy} combinations.
\begin{enumerate}[]
  
\item \case[1]{$\exists q,q' \in \Msingle : L(\Srest) \leq 1$.}
\label{case:lrest_leq_1}

\begin{addmargin}[2em]{0em}
Adding $\ieight, \ieight'$ to $\Seight$, $q$ to $\Meight$, $\Srest$ to $\Snice$ and $q'$ to $\Mnice$ is enough: This preserves $\Lnice(\Snice) \leq \abs{\Mnice}$.
We do this for all pairs of $q, q'$ where this is applicable.
Afterwards, either $\abs{\Msingle} \leq 1$, or it holds for all pairs $q,q' \in \Msingle : L(\Srest) > 1$.
\end{addmargin}

\item \case[2]{$\exists q,q' \in \Msingle, i, i' \in \typefive \cup \typesix : \solopt(i) = q \wedge \solopt(i') = q'$.}
\label{case:lrest_nicetypes}

\begin{addmargin}[2em]{0em}
We may also simply add $\ieight, \ieight'$ to $\Seight$ and $q$ to $\Meight$: By definition of nice instances, the instance $I'$ consisting of all job pieces in $\Srest$ and $q'$ is itself a nice instance, for which $\mnice(I') = 1$.
As $\Lnice(\Inice) \leq \abs{\Mnice}$, $\mnice(\Inice) \leq \abs{\Mnice}$ by \autoref{lemma:lnice_implies_mnice}.
Thus, $\mnice(\Inice \cup I') \leq \abs{\Mnice} + 1$ by definition of $\mnice$, and \autoref{preemptive:simple},~(\ref{preemptive:simple:decision:T_check_true}) holds for $\Inice$ after adding $\Srest$ to $\Inice$ and $q'$ to $\Mnice$.
We again do this for all pairs of $\ieight, \ieight'$ where this is applicable.
This schedules at least all but one such machine $q \in \Msingle$.
We add $q$ and all setup times and job pieces scheduled on it in $\solopt$ to $\Ieightrem$.
\end{addmargin}

\item \case[3]{$\exists q,q' \in \Msingle, i,i' \in \typeseven : s_i + P(C_i) \geq \sfrac12, s_{i'} + P(C_{i'}) \geq \sfrac12$} \\ \underline{and $i,i'$ are fully scheduled on $q,q'$, respectively.}
\label{case:lrest_big7classes}

\begin{addmargin}[2em]{0em}
By definition of $\solopt$, we know that $s_i + P(C_i) \leq \sfrac59$ and $s_{i'} + P(C_{i'}) \leq \sfrac59$.
We fully schedule $i$ and $i'$ on $q'$ (with completion time at most $\sfrac{10}{9}$), and put $\Srest \setminus \{i,i'\}$ into $\Seight$, $q$ into $\Meight$.
Note that, by definition of $i,i'$, $L(\Srest \setminus \{i,i'\}) \leq \sfrac19$.
By doing this for all such pairs where this is applicable, there is at most one such $q \in \Msingle$ with class $i$ as defined above remaining afterwards.
We add $\ieight, i$ and $\Srest(q)$ to $\Seight$ and $q$ to $\Meight$.
Note that $L(i) + L(\ieight) \leq 1$.
\end{addmargin}
\end{enumerate}

After this, it hold for all pairs of machines $q,q' \in \Msingle$ that only setup times and job pieces of $\typeseven$ are in $\Srest$, and $L(\Srest) > 1$.
Furthermore, $L(\ieight) + L(\iten) > \sfrac43$, for all $\ieight \in \typeeight \cup \Seight, \iten \in \typeten \cup \Seight$: Otherwise, our matching of $\typeeight$ and $\typeten$ classes would not have been maximal.
Let $\iten$ be the class with smallest load over all classes in $\typeten \cup \Seight$, $L(\iten) = \sfrac89 - Y'$ for some $Y' \in \Z$, and $Y = \max(0,Y')$.
This definition implies $0 \leq Y \leq \sfrac{1}{18}$.
Then it holds for all $\ieight \in \typeeight \cap \Seight$ that $L(\ieight) \geq \sfrac49 + Y$: Otherwise, $L(\iten) + L(\iten) \leq \sfrac43$.
Note that, as $L(\ieight) + L(\ieight') \geq \sfrac89 + 2 Y$ for all such classes $\ieight, \ieight'$, and thus, $L(\Srest) \leq \sfrac{10}{9} - 2 Y$.

Let $\Srest(q), \Srest(q') \subseteq \Srest$ be the subsets of $\Srest$ originally scheduled on $q$ and $q'$, and $L(\Srest(q)) = \sfrac{1}{2} + x$, $L(\Srest(q')) = \sfrac{1}{2} + x'$, where $\sfrac{-1}{18} \leq x, x' \leq \sfrac{1}{18} - Y$ by definition of $q,q'$.
By the steps before, we know that $x + x' > 0$ for all such pairs.
In the following, we will always add $q'$ to $\Mnice$, and add $q$ to $\Meight$ or directly schedule some classes on it; to preserve $\Lnice(\Inice) \leq \abs{\Mnice}$, it is then enough to add at most $(L(\Srest(q)) - \max(0,x)) + (L(\Srest(q')) - \max(0,x'))$ to $\Snice$.
We show how to do this by a full case distinction on the types of classes and jobs in $\Srest(q)$ and $\Srest(q')$.
If $\abs{\Msingle}$ is not even, we take remove a machine $q$ from $\Msingle$, and add it, as well as all setup times and job pieces scheduled on it in $\solopt$, to $\Ieightrem$.
Now, while $\abs{\Msingle} \neq \emptyset$, let $q,q' \in \Msingle$ again be a pair of machines.
We show the procedure for $\Srest(q)$ and $x$, doing this for $\Srest(q')$ and $x'$ follows symmetrically.
Adding jobs from $\Srest$ to $\Seight$ or scheduling them directly on $q$ always implies the removal of these jobs from $\Srest$.
In some of the following cases, we remove enough load from either $\Srest(q)$ or $\Srest(q')$ such that the process need not be executed for $q'$ or $q$.
If this occurs for either $q$ or $q'$, we assume wlog. that it always occurs for $q$, \ie the machine we consider first, and simply add all $\Srest(q')$ to $\Snice$.
The process as given by the case distinction is only executed for $q$ (resp., $q'$) if $x > 0$ (resp., $x' > 0$), as otherwise, $L(\Srest(q))$ (resp., $L(\Srest(q'))$) is small enough already for our purposes.
We write $\exists i \in \Srest(q)$ iff there are some job pieces of class $i$ in $\Srest(q)$.
\begin{enumerate}[]
\setcounter{enumi}{3}
\item \case[4]{$\exists i \in \Srest(q) : s_i > x$ $\wedge$ $s_i$ appears more than once in $\Srest(q)$.}
\label{case:morethan1si}

\begin{addmargin}[2em]{0em}
Adding $\Srest(q)\setminus s_i$ to $\Snice$ is enough, as $L(\Srest(q)\setminus s_i) = (\sfrac12 + x) - s_i \leq \sfrac12$, and only one $s_i$ is counted in $\Lnice(\Inice)$.
\end{addmargin}

\item \case[5]{$\exists i \in \Srest(q) : s_i > x$ $\wedge$ $i$ is not fully scheduled on $q$.}
\label{case:notfully}

	\begin{enumerate}[]

    \item \case[5.1]{$s_i \leq \sfrac{1}{6}$.}
    \label{case:notfully_leq16}
    
    \begin{addmargin}[2em]{0em}
    If $s_i \in \Snice$ already before the beginning of this process, we simply add $\Srest(q)\setminus s_i$ to $\Snice$, maintaining $\Lnice(\Inice) \leq \abs{\Mnice}$, and add $\ieight,\ieight'$ to $\Seight$.
    Thus, in the following assume otherwise.
    Let $\Msingle^i \subseteq \Msingle$ be the subset of machines in $\Msingle$ with job pieces of $i$ scheduled on them.
    Let $P' = s_i$ before the first machine in $\Msingle^i$ is considered in this process.
    
    We add $J', \ieight, \ieight'$ to $\Seight$, for some set of job pieces $J' \in C_i, J' \in \Srest(q)$ with $P(J') = \min(\sum_{j \in C_i, \solopt(j) = q} p_j,P')$.
    If $s_i \notin \Seight$, we additionally add $s_i$ to $\Seight$.
    If $P(J') < P'$ and $s_i \notin \Snice$, we add $\Srest(q)\setminus J'$ to $\Snice$; otherwise, we add  $\Srest(q)\setminus \{J',s_i\}$ to $\Snice$.
    We set $P' = P' - P(J')$.
    
    Essentially, we make sure that only one $s_i$ is included in $\Seight$ and $\Snice$, respectively.
    Before adding $s_i$ to $\Snice$, we also make sure to add at least $s_i$ processing time of jobs in $C_i$ to $\Seight$ instead of $\Snice$.
    We show that this this is enough to keep the overall load added to $\Snice$ small enough such that $\Lnice(\Inice) \leq \abs{\Mnice}$ is preserved.
    Let $p'_j$ be the processing time of a job $j$ over all machines $q_i \in \Msingle^i$ in $\solopt$.
    If $\sum_{j \in C_i} p'_j \geq s_i$, the load added to $\Snice$ over all machines in $\Msingle^i$ can be expressed as
    
    $$\sum_{q_i \in \Msingle^i} (\Lrest(q_i) - s_i) + s_i - s_i \leq \sum_{q_i \in \Msingle^i} (\sfrac59 - x) \leq \sfrac{\abs{\Msingle^i}}{2} \enspace ,$$
    
    as there is at most one setup time $s_i$ added to $\Snice$ over all machines in $\Msingle^i$, but also at least $s_i$ processing time of $\sum_{j \in C_i} p'_j$ not added to $\Snice$ in the process.
     If $\sum_{j \in C_i} p'_j < s_i$, the load added to $\Snice$ over all machines in $\Msingle$ can instead be expressed as $\sum_{q_i \in \Msingle^i} (\Lrest(q_i) - s_i) \leq \sfrac{\abs{\Msingle^i}}{2}$, as $s_i$ is always in $\Srest(q_i)$ for all $q_i$, but never added to $\Snice$.
     This maintains our goal of adding at most $(L(\Srest(q_i)) - \max(0,x)) \leq \sfrac12$ load per machine $q_i \in \Msingle^i$ to $\Snice$.
     
    Now, we look at the average load added to $\Seight$ over all such machines.
    As $s_i \notin \Snice$ before the beginning of this process, and $i$ is not fully scheduled on $q$, there are two possibilities:
   	There exists some $q_i \in \Mdual \cup \Mten$ such that $i$ is partly scheduled on $q_i$ in $\solopt$, or not.
   	In the former case, at most $s_i$ load is added to $\Seight$ per machine $q$.
   	In the latter case, there is at most $2 s_i$ load added to $\Seight$, but since $i$ is not fully scheduled on $q$, but neither on machines in $\Mnice$ or $\Mdual \cup \Mten$, it must be scheduled on another machine in $\Msingle$.
   	Thus, we add at most $s_i \leq \sfrac16$ load to $\Seight$ per machine.
   	As this added load always included a setup time $s_i$, we add at most $s_i$ processing time of $i$ to $\Seight$ here.
   	If, in fact, $s_i > \sfrac19$, we already reduce the load of $\Srest$ by at least $x + x'$ per machine.
	Thus, the process is not executed for $q'$.

	\end{addmargin}
	
	\item \case[5.2]{$s_i> \sfrac{1}{6}$.}
	\label{case:notfully_g16}
	
	\begin{addmargin}[2em]{0em}
	Then all other job pieces of $i$ scheduled on some other machines in $\solopt$ are scheduled on other machines in $\Msingle$, are already included in $\Snice$, or both: Due to the high setup time, such job pieces cannot be scheduled on machines of $\Mdual \cup \Mten$ in $\solopt$.
	If some job piece is already in $\Snice$, then so is $s_i$, and we simply add $\Srest(q) \setminus s_i$ into $\Snice$, with the load of this added set being at most $\sfrac59 - s_i < \sfrac12$.
	If this is not the case, then there are some other job pieces of class $i$ scheduled on at least one other machine in $\Msingle$.
	Let $\Msingle^i \subseteq \Msingle$ be the set of these machines.
	We add $\Srest(q_i) \setminus s_i$ to $\Snice$, for all $q_i \in \Msingle^i$, as well as $s_i$ only once.
	This implies that the load put into $\Snice$ over all $\abs{\Msingle^i} \geq 2$ of these machines is at most
	\begin{equation*}
	\begin{aligned}
	s_i + \abs{\Msingle^i} (\sfrac{5}{9} - s_i) &= \sfrac{\abs{\Msingle^i}}{2} + \sfrac{\abs{\Msingle^i}}{18} - (\abs{\Msingle^i}-1) s_i \\
	&\leq \sfrac{\abs{\Msingle^i}}{2} \enspace ,
	\end{aligned}
	\end{equation*}
	where the last inequality follows from the fact that $\sfrac{1}{6} < s_i$ and $\abs{\Msingle^i} \geq 2$.
	This maintains our goal of adding at most $(L(\Srest(q_i)) - \max(0,x)) \leq \sfrac12$ load per machines $q_i \in \Msingle^i$ to $\Snice$.
	There is no load added to $\Seight$ in this case.

	\end{addmargin}

   \end{enumerate}
    
\item \case[6]{$\exists i \in \Srest(q) : s_i > x \wedge i$ is fully scheduled on $q$ $\vee$ $\nexists i \in \Srest(q) :$} \\ \underline{$s_i > x$.}
\label{case:fully}

\begin{addmargin}[2em]{0em}
We know that, if such an $i$ exists, that $s_i + P(C_i) < \sfrac12$, as otherwise, $q$ would have been considered already in \caseref{case:lrest_big7classes}{3}.
\end{addmargin}

\begin{enumerate}[]

    \item \case[6.1]{$s_i + P(C_i) \leq \sfrac16$.}
    \label{case:fully_l19}
    
    \begin{addmargin}[2em]{0em}
    Then also $s_i \leq \sfrac16$.
	We add $s_i + P(C_i)$, \ie at most $\sfrac16$ load, to $\Seight$.
	If, in fact, $s_i + P(C_i) > \sfrac19$, we already reduce the load of $\Srest$ by at least $x + x'$ per machine, and the process is not executed for $q'$.
	Otherwise, $s_i \leq \sfrac19$.
	This reduces the load of $\Srest(q)$ by at least $s_i + P(C_i) > x$, which we then can simply add to $\Snice$.
	\end{addmargin}

    \item \case[6.2]{
    $\sfrac16 < s_i + P(C_i) \leq \sfrac13$.}
    \label{case:fully_leq13}
    
    \begin{addmargin}[2em]{0em}
    We add $\Srest(q) \setminus \{s_i,C_i\}$ to $\Snice$.
    If $s_i \leq \sfrac19$ and $s_i + P(C_i) \leq \sfrac29$, we add $s_i + P(C_i)$ to $\Seight$.
    Otherwise, we directly schedule $s_i + P(C_i)$ on $q$, starting at $0$, and both $\ieight$ and $\ieight'$ above, starting at $\sfrac13$.
     This maintains $\Lnice(\Inice) \leq \abs{\Mnice}$: The load of $\Snice$ is overall increased by at most $L(\Srest) - \sfrac16 \leq 1$, while $q'$ is added to $\Mnice$.
    In this case, as $s_i + P(C_i) > \sfrac16 > x+x'$, the process is not executed for $q'$.
    \begin{tikzpicture}[xscale=0.5,yscale=2]
		\tikzmath{\m = 2;}
        \hlines \m {
            {4/3}/$\sfrac43$//,
            1/$1$/solid/,
            {2/3}/$\sfrac23$//,
            {1/3}/$\sfrac13$//,
            {1/9}/$\sfrac19$//,
            0/$0$/solid/
        };
        \schedule up {
            {
                {2/9}//{preaction={fill,gray},pattern=north east lines}/,
            	{1/18}//{preaction={fill,lightgray},pattern=north east lines}/,
                {1/18}//opacity=0/,
                {4/9}/8/fill=lightgray/\encircleinsched,
                {1/2}/8/fill=lightgray/\encircleinsched
             }/1/
        };
        \node at (1.8,1/2) {\dots};     %
    \end{tikzpicture}
	\end{addmargin}

    \item  \case[6.3]{$\sfrac12 > s_i + P(C_i) > \sfrac13$ $\vee$ $\nexists i \in \Srest(q), s_i > x$.}
    \label{case:fully_tiny}

\begin{addmargin}[2em]{0em}    
    If such an $i$ exists, it is the only class scheduled on $q$ in $\solopt$ with $s_i > x$, as only one class of processing time $> \sfrac13$ can be fully scheduled on $q$ besides $\ieight$.
    Then $i$ is also scheduled on $q$ without interruption:
    Otherwise, one would need at least two setups of size $s_i > x$, which would have been handled by \caseref{case:morethan1si}{4}.
    Therefore, the setup times and job pieces in $\Srest \setminus i$ all belong to some classes with setup times $< x$, and are scheduled below both $\ieight$ and $i$, between these two, or above both in $\solopt$, as pictured below.

    
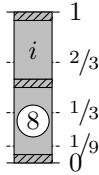
\tikz[xscale=0.5,yscale=2]{
		\tikzmath{\m = 1;}
        \hlines \m {
            1/$1$/solid/,
            {2/3}/$\sfrac23$//,
            {1/3}/$\sfrac13$//,
            {1/9}/$\sfrac19$//,
            0/$0$/solid/
        };
        \schedule up {
            {
            	{1/18}//{preaction={fill,lightgray},pattern=north east lines}/,
                {4/9}/8/fill=lightgray/\encircleinsched,
                {1/18}//{preaction={fill,lightgray},pattern=north east lines}/,
                {1/3+1/18}/$i$/fill=lightgray/,
                {1/18}//{preaction={fill,lightgray},pattern=north east lines}/
             }/1/
        };
     }

     Since $s_i + P(C_i) < \sfrac12$, the total load of $L(\Srest) - (s_i + P(C_i))$ on $q$ is greater than $x$.
     We now take at least $x$ load of classes with $s_i \leq x$ from $q$ and add it to $\Seight$,
     taking only job pieces that are scheduled in the intervals $[\sfrac{8}{9} + 2Y,1]$, $[\sfrac{4}{9} + Y,\sfrac{5}{9} - Y]$, and $[0,\sfrac{1}{9} - 2Y]$ on $q$ in $\solopt$.
    We know that, in $\solopt$, there is at least $x$ load in each interval of size $2(\sfrac{1}{18} - Y)$, as there is at most $1 - (\sfrac49 + Y) - (\sfrac12 + x) = \sfrac{1}{18}-Y-x \leq \sfrac{1}{18}-Y$ free space on each machine and $x \leq \sfrac{1}{18} - Y$ by definition.
	In fact, over these three intervals together, there is at least $x$ load of classes with setup times $\leq x$ scheduled in $\solopt$: The only classes on $q$ with setup time $> x$ are $\ieight$ and $i$, if $i$ exists.
	As they are both scheduled on $q$ without interruption, and their combined load is less than $1-x$, no matter where they are scheduled on $q$, \ie either one at the top and at the bottom or both in the middle, there is always at least $x$ load of classes with setup time $\leq x$ over the three intervals.
	To achieve this, we execute the following procedure:
\begin{algorithmic}[1]
\Procedure{RemoveTinyJobs}{$x,q$}
	\State $ret = \emptyset$ 
	\ForAll{$\mathcal{I} \in [\sfrac{8}{9}+2Y,1],[\sfrac{4}{9}+Y,\sfrac{5}{9}-Y],[0,\sfrac{1}{9}-2Y]$}
		\If{$x > 0$}
			\State Let $C'_{x} := \{C'_1, \dots, C'_c\}$, where $C'_k$ is defined to be the set of job parts of jobs in $C_k$ that are scheduled in $\mathcal{I}$ on $q$ if $s_k \leq x$, and $C'_k = \emptyset$, else, for all $k \in \iv c$.
			For simplicity reasons, here, let $s_k := 0$ if $C'_k = \emptyset$.
			\If{$\sum_{k \in C'_{x}} (s_k + P(C'_k)) \leq x$}
				\State $ret = ret \cup \bigcup_{k \in  C'_{x}} s_k \cup C'_{x}$
				\State $x = x - \sum_{k \in C'_{x}} (s_k + P(C'_k))$
			\Else
				\State $k=1$
				\While{$x>0$}
					\If{$s_k + P(C'_k) \leq 2x$}
						\State $ret = ret \cup s_k \cup C'_k$
						\State $x = x - s_k - P(C'_k)$
						\State $k = k+1$
					\Else
						\State Let $J^x \subseteq C'_k$ be a subset of job pieces with proc. time exactly $x$.
$J^x$ always exists in $\mathcal{I}$ due to the fact that $s_k \leq x$ and $x \leq \sfrac{1}{18} - Y$.
						\State $ret = ret \cup s_k \cup J^x$
						\State $x = x - s_k - P(J^x)$
					\EndIf
				\EndWhile
			\EndIf
		\EndIf
	\EndFor
	\Return $ret$
\EndProcedure
\end{algorithmic}
	 By definition of this procedure, $ret$ only included jobs from the three intervals, and $x \leq L(ret) \leq 2x$.
	 Thus, we add at most $2x \leq \sfrac19$ load to $\Seight$, while reducing the load of $\Srest(q)$ by at least $x$, making it safe to add the remainder of $\Srest(q)$ to $\Snice$ while preserving $\Lnice(\Inice) \leq \abs{\Mnice}$.
\end{addmargin}
\end{enumerate}
\end{enumerate}

After the process above, only the setup times and job pieces in $\Seight$ are not yet scheduled on machines with makespan at most $\sfrac43$.
We know aim to schedule these on the machines in $\Meight$ with makespan at most $\sfrac43$, thus obtaining a $\sfrac43$-approximation for the whole instance.
First note that, by construction, there is at least one machine in $\Meight$ for each class of $\typeten$ in $\Seight$, and at least one machine for each two classes of $\typeeight$ in this set, as well as one machine for classes $\ieight$ and $i$ added to $\Seight$ in \caseref{case:lrest_big7classes}{3}.
We start by scheduling each such class of $\typeten$ on one machine of $\Meight$ starting at $\sfrac13$, and each pair of such $\typeeight$ classes as well as $\ieight$ and $i$ on one machine of $\Meight$ one after another, also starting at $\sfrac13$.
We assume that now, each machine in $\Meight$ either has one $\typeten$ class or two $\typeeight$ classes scheduled on it, always starting at $\sfrac13$.

Afterwards, all non scheduled job pieces in $\Seight$ belong to classes $i \in \typeseven$.
We now want to place these job pieces on the machines of $\Meight$, between $0$ and $\sfrac13$.
If we manage to do this in a feasible way, we are done: All other classes of $\typeseven$ are part of $\Inice$, where all job pieces of $\typeseven$ classes are scheduled above $\sfrac13$ by definition of the algorithm given in \autoref{preemptive:simple}.
Note that the only times setup times and job pieces are added to $\Seight$ are at the start of this section, where all setup times and job pieces machines in $\Mdual \cup \Mten$ are added to $\Seight$, and in cases \caseref{case:lrest_big7classes}{3}, \caseref{case:notfully_leq16}{5.1}, and \caseref{case:fully}{6}.
As the total load of setup times and processing times that are added to $\Seight$ is at most $\sfrac29$ per machine added to $\Meight$, it holds that $L(\Seight) \leq \sfrac29 \abs{\Meight}$.

To schedule the job pieces in $\Seight$, we use \wrapdyn{}.
For this, we construct a wrap sequence $Q$ from $\Seight$ and a respective wrap template $\omega$ on the machines in $\Meight$.
To show feasibility of \wrapdyn{} for $Q$ and $\omega$, we prove that the prerequisites of both \autoref{prop:wrapdyn:space} and \autoref{prop:wrapdyn:para} are fulfilled.

Let $Q : = (s_{i_1},j_1^1,\dots,j_{n_1}^1,s_{i_2},j_1^2,\dots,j_{n_2}^2,\quad\dots\quad,s_{i_k},j_1^k,\dots,j_{n_k}^k)$ be the wrap sequence constructed by taking all job pieces in $\Seight$ grouped by their class, and adding the setup time of their respective class to the sequence, and let $\omega = (\omega_1, \dots, \omega_{\abs{\Meight}})$, with $\omega_r = (r,0,\sfrac13), \forall 1 \leq r \leq \abs{\Meight}$.
Since there is at least one setup time per class of every job piece in $\Seight$ by construction of the set, and only one per such class in $Q$, $L(Q) \leq L(\Seight)$.

We first show that the requirement of \autoref{prop:wrapdyn:space} is fulfilled. Since $b_r - a_r = \sfrac13 > \max_{r \in \abs{\omega}} s_r$ is fulfilled for all $r \in [\abs{\omega}]$ by virtue of all job pieces in $Q$ being part of $\typeseven$ classes, it is only left to show that $L(Q) \leq S(\omega) - \sum_{r=2}^{\abs{\omega}} s_r$.
First note that, for all job pieces $j \in \Seight$ with $s_j > \sfrac19$, all such $j$ are job pieces that were previously scheduled on machines in $\Mten$, or that were added in \caseref{case:notfully_leq16}{5.1} or \caseref{case:fully_l19}{6.1}: In all other cases, all job pieces added to $\Seight$ belong to classes $i$ with $s_i \leq \sfrac19$.
Simultaneously, the maximum setup time $s_j$ of such $j$, as well as the maximum load added to $\Seight$ per machine, is $\sfrac16$, either by the definition of $\typeten$, as shown in \caseref{case:notfully_leq16}{5.1}, or by definition of \caseref{case:fully_l19}{6.1}.
In all other cases, \ie for all $j \in \Seight$ with $s_j \leq \sfrac19$, the maximum load added to $\Seight$ per machine is $\sfrac29$.
We again define $p'_j$ to be total processing time of all job pieces of a job $j$ included in $\Seight$.
Let $L_1 = \sum_{j \in \Seight, s_j > \sfrac19} p'_j + \sum_{i \in \Seight, s_i > \sfrac19} s_i$ and $L_2 = \sum_{j \in \Seight, s_j \leq \sfrac19} p'_j + \sum_{i \in \Seight, s_i \leq \sfrac19} s_i$.
By this definition, $L_1 + L_2 = L(Q)$.
Then, by the maximum load added to $\Seight$ per machine in $\Meight$, there exists some $r' \in \abs{\omega}$ with $\sfrac16 r' + \sfrac29 (\abs{\omega} - r') \geq \ceil*{L_1 + L_2}$, such that $L_1$ can be fully scheduled on machines $1, \dots, r'$, and $L_2$ can be fully scheduled on machines $r', \dots, \abs{\omega}$.
This implies that $s_r \leq \sfrac16$, for the additional setup times $s_r$ on machines $1, \dots, r'$, and $s_r \leq \sfrac19$ for the additional setup times $s_r$ on machines $r'+1, \dots, \abs{\omega}$.
Thus,
\begin{align*}
\label{eq:eightrem_wrapload}
S(\omega) - \sum_{r=2}^{\abs{\omega}} s_r &\geq \sfrac13 \abs{\omega} - \sum_{r=2}^{r'} \sfrac16 - \sum_{r=r'+1}^{\abs{\omega}} \sfrac19 \\
&= \sfrac13 r' + \sfrac13 (\abs{\omega} - r') - \sfrac16 r' - \sfrac19 (\abs{\omega} - r') \\
&\geq \sfrac16 (r'-1) + \sfrac29 (\abs{\omega} - r') \geq L_1 + L_2 - \sfrac16 \\
&= L(Q) - \sfrac16 \enspace ,
\end{align*}
and the claim follows.

We now show that the requirement of \autoref{prop:wrapdyn:para} is fulfilled.
It holds for all $\omega_r, \omega_{r+1}$ that $u_r < u_{r+1}, b_r = b_{r+1}, a_r = a_{r+1}, \forall r \in [1,\abs{\omega}-1]$.
Thus, it is only left to show that
$\max_{\ell \in \iv k, y \in n_{\ell}} s_{i_{\ell}} + j^{\ell}_y \leq b_r - a_r, \forall r \in [\abs{\omega}]$.
To prove this, it is enough to show that for every job $j \in \Seight$, $s_j + p'_j \leq \sfrac13$, where $p'_j$ is the processing time of $j$ included in $\Seight$.
We first look at all jobs $j \in \Seight$ that belong to some class $i$ with $s_i > \sfrac19$.
As argued before, all such jobs were previously scheduled on machines in $\Mten$ in $\solopt$, or were added in \caseref{case:notfully_leq16}{5.1} or \caseref{case:fully_l19}{6.1}.
In \caseref{case:fully_l19}{6.1}, only full classes $i$ with $s_i + P(C_i) \leq \sfrac16$ are added to $\Seight$.
Thus, no job piece of a job of this class is added in any other step, and $s_i + p'_j \leq \sfrac16$.
There is at most $\sfrac13-2s_i$ processing time of a single job $j$ of $C_i$ on machines in $\Mten$ by definition of \typeten. 
As there is at most $s_i$ processing time of jobs in $C_i$ to $\Seight$ in \caseref{case:notfully_leq16}{5.1}, it holds for each such job $j$ that $s_i + p'_j \leq s_i + \sfrac13-2s_i+ s_i \leq \sfrac13$.

Now, consider all jobs $j \in \Seight$ that belong to some class $i$ with $s_i \leq \sfrac19$.
Pieces of such jobs may have been previously scheduled on machines of $\Mdual \cup \Mten$ in $\solopt$, or added to $\Seight$ in cases \caseref{case:lrest_big7classes}{3}, \caseref{case:notfully_leq16}{5.1}, or \caseref{case:fully}{6}.
Any job piece added in \caseref{case:fully_l19}{6.1} or \caseref{case:fully_leq13}{6.2} belongs to a class $i$ that was fully scheduled on a machine $q \in \Msingle$ in $\solopt$.
This implies that no job piece of this class was scheduled on any other machine in $\solopt$.
Thus, $s_i + p'_j \leq \sfrac13$ by definition of these cases.

We now analyze the total processing time of job pieces of $j$ in $\Seight$ that have been added to this set in any or all of the other cases.
For this, we look at the total size of the intervals where job pieces are taken from $\solopt$ and added to $\Seight$.
As $\solopt$ is a feasible solution with no parallelization, the total processing time of $j$ added to $\Seight$ over all these intervals in all these cases is at most the total non-overlapping size of these intervals minus one required setup time per interval.
On machines of $\Mten$, pieces of $j$ can only be scheduled in the intervals $[\sfrac89-Y,1]$ and $[0,\sfrac19+Y]$ by definition of $Y$.
On machines of $\Mdual$
, as both $\ieight$ and $\ieight'$ are scheduled without interruption, $j$ can only be scheduled in the intervals $[\sfrac89+2Y,1]$, $[\sfrac49+Y,\sfrac59-Y]$ and $[0,\sfrac19-2Y]$ on such machines.
Similarly, for machines considered in \caseref{case:lrest_big7classes}{3},
pieces of $j$ can only be scheduled in intervals $[\sfrac89+Y,1]$, $[\sfrac49+Y,\sfrac59-Y]$ and $[0,\sfrac19-Y]$ on such machines.
By definition of \caseref{case:fully_tiny}{6.3}, all job pieces of $j$ in $\Seight$ were scheduled only in the intervals $[\sfrac{8}{9} + 2Y,1]$, $[\sfrac{4}{9} + Y,\sfrac{5}{9} - Y]$, and $[0,\sfrac{1}{9} - 2Y]$.
This situation is visualized in \autoref{fig:single8_sj_int}.

\begin{figure}
	\centering
	\tikz[xscale=0.5,yscale=2]{
		\tikzmath{\m = 6;\smalleps = 1/18;}
        \hlines \m {
            1/$1$/solid/,
            {5/6}/$\sfrac56$//,
            {2/3}/$\sfrac23$//,
            {1/2}/$\sfrac12$//,
            {1/3}/$\sfrac13$//,
            {1/6}/$\sfrac16$//,
            0/$0$/solid/
        };
        \schedule up {
            {
                {4/9}/8/fill=lightgray/\encircleinsched,
                {4/9+\smalleps}//{preaction={fill,lightgray},pattern=north east lines}/
             }/1/,
             {
                {1-1/9}/10/fill=lightgray/\encircleinsched,
                {1/9}//{preaction={fill,lightgray},pattern=north east lines}/
             }/1/,
             {
                {1/9-\smalleps}//opacity=0/,
                {4/9+\smalleps}//{preaction={fill,lightgray},pattern=north east lines}/,
                {4/9}/8/fill=lightgray/\encircleinsched
             }/1/,
            {
                {1/9}//{preaction={fill,lightgray},pattern=north east lines}/,
                {1-1/9}/10/fill=lightgray/\encircleinsched
            }/1/,
            {
            	{2/9+\smalleps}//{preaction={fill,lightgray},pattern=north east lines}/,
                {4/9}/8/fill=lightgray/\encircleinsched,
                {2/9}//{preaction={fill,lightgray},pattern=north east lines}/
             }/1/,
            {
                {4/9}/8/fill=lightgray/\encircleinsched,
                {1/9}//{preaction={fill,lightgray},pattern=north east lines}/,
                {4/9}/8/fill=lightgray/\encircleinsched
             }/1/
        }; 	}
	\caption{The position of \typeeight and \typeten classes in $\solopt$.}
	\label{fig:single8_sj_int}
\end{figure}
	
The total non-overlapping size of these intervals, minus one required setup time $s_i$ per interval, then is
$$2 (\sfrac19 + Y - s_i)+ (\sfrac19 - 2Y - s_i) = \sfrac13 - 3 s_i$$
The only time a job piece of $j$ may have been added to $\Seight$ outside these intervals is in \caseref{case:notfully_leq16}{5.1}.
The total processing time of a single class, and thus also a single job, added to $\Seight$ in \caseref{case:notfully_leq16}{5.1} is at most $s_i$.
Thus, $p'_j \leq \sfrac13 - 2 s_i$ and $s_i + p'_j \leq \sfrac13 - s_i$.

The only thing left to do now is to deal with $\Ieightrem$.
In total, we have added setup times, job pieces and machines to $\Ieightrem$ at most three times:
\begin{enumerate}
\item When pairing machine of $\Msingle \cup \Mdual$ which have some part of a nice class in $\typeone, \dots, \typesix$ scheduled on them, where we add
no machines and
a class $\ieight^1 \in \typeeight$
to $\Ieightrem$.
\item When pairing machines of $\Msingle$ that have class $i_{5,6} \in \typefive \cup \typesix$ fully scheduled on them, where we add one machine $q_2$, a class $\ieight^2 \in \typeeight$ and $\Srest(q_2)$ to $\Ieightrem$.
\item When pairing the rest of the machines of $\Msingle$, where we add one machine $q_3$, a class $\ieight^3 \in \typeeight$, and $\Srest(q_3)$ to $\Ieightrem$.
\end{enumerate}

We also know that there is currently no class of $\typeeight$ in $\Snice$ by all previous steps of our algorithm, but a nice instance allows for at most one class of $\typeeight$ by \autoref{def:nice_instance}.
The number of machines in $\Ieightrem$ is thus either $0$, $1$, or $2$.

If the number of machines is $0$ or $2$, neither or both $q_2 \in \Ieightrem$ and $q_3 \in \Ieightrem$.
In the latter case, we schedule $\ieight^2$ and $\ieight^3$ together on $q_2$, starting at $\sfrac13$.
Then, the sub-instance $I'$ induced by $\Srest(q_2) \cup \Srest(q_3)$ and $q_3$ has $\mnice(I') = 1$.
As $\Lnice(\Inice) \leq \abs{\Mnice}$, $\mnice(\Inice) \leq \abs{\Mnice}$ by \autoref{lemma:lnice_implies_mnice}.
Thus, $\mnice(\Inice \cup I') \leq \abs{\Mnice} + 1$ by definition of $\mnice$, and \autoref{preemptive:simple},~(\ref{preemptive:simple:decision:T_check_true}) holds for $\Inice$ after adding  $\Srest(q_2) \cup \Srest(q_3)$ to $\Inice$ and $q_2$ to $\Mnice$.
Now, either there are no more setup times and job pieces in $\Ieightrem$, and we are done, or only $\ieight^1$ is in $\Ieightrem$.
Since, when adding the machine $q_1$ on which $\ieight^1$ was scheduled in $\solopt$, we have increased $\abs{\Mnice}$ by one, but added only at most $1 - L(\ieight^1) - \sfrac13 \geq \sfrac29$ load to $\Snice$, implying that $\sfrac79$ load can be added to $\Snice$ without violating $\Lnice(\Inice) \leq \abs{\Mnice}$.
We can thus simply add $\ieight^1$ to $\Snice$ as the single class of \typeeight in $\Inice$.

If the number of machines is $1$, either machine $q_2 \in \Ieightrem$ or $q_3 \in \Ieightrem$.
If now $\ieight^1 \notin \Ieightrem$, we simply add $q \in \{q_2,q_3\}$ to $\Mnice$, and all setup times and job pieces that were scheduled on $q$ in $\solopt$ to $\Snice$, keeping the instance nice with exactly one class of $\typeeight$.
If instead $\ieight^1 \in \Ieightrem$, we schedule both classes of \typeeight in $\Ieightrem$ on $q$, and add all setup times and job pieces previously scheduled on $q$ but the \typeeight class to $\Snice$.
As the total load of these setup times and job pieces is at most $\sfrac59$, but we can add $\sfrac79$ to $\Snice$ without violating $\Lnice(\Inice) \leq \abs{\Mnice}$ if $\ieight^1 \in \Ieightrem$, the resulting schedule is feasible.

The resulting schedule is now a feasible $\sfrac43$-approximation of the original instance.
This then, in total, shows that for every optimal solution $\solopt$ of an instance with makespan $T$ and more classes of $\typeeight$ than $\typenine$, there exists a feasible schedule with makespan at most $\sfrac43 T$, where
\begin{itemize}
\item all classes of $\{\typeone, \dots, \typesix\}$ and some partial classes of $\typeseven$ are part of a nice instance
\item all classes of \typenine are scheduled together with classes of \typeeight  with no other setup times or job pieces on one machine each
\item all pairs of classes of \typeeight and \typeten with combined load less than $\sfrac43 T$ are scheduled together with no other setup times or job pieces on one machine each
\item all other but at most two \typeeight classes are scheduled in pairs on a machine each with starting time $\sfrac13 T$
\item of those at most two \typeeight classes, one is included in the nice instance, and the other is scheduled together with a \typeseven class $i$ with load  $\sfrac59 T \geq s_i + P(C_i) \geq \sfrac12 T$ on one machine  with starting time $\sfrac13 T$
\item all other \typeten classes are scheduled on a machine each with starting time $\sfrac13 T$
\item all remaining \typeseven classes $i$ not included in the nice instance with load $\sfrac59 T \geq s_i + P(C_i) \geq \sfrac12 T$ are scheduled in pairs on a machine each
\item all other remaining \typeseven classes $i$ have $s_i \leq \sfrac16 T$ and are scheduled below $\sfrac13 T$ together with classes of \typeeight and \typeten above, with at most $\sfrac29 T$ processing time per machine for all classes such \typeseven classes $i$ with $s_i \leq \sfrac19 T$, and at most $\sfrac16 T$ processing time per machine for all such \typeseven classes $i'$ with $\sfrac16 T \geq s_{i'} > \sfrac19 T$
\end{itemize}

\subsection{Classes of Type Nine Remain}
\label{sec:nine_rem}
Now, we consider the case where, after our matching of \typeeight and \typenine classes, the only machines with non-nice types of classes not yet modified contain full classes of \typenine and \typeten.
Let $\Mnt$ be the subset of machines with these classes scheduled on them in $\solopt$.
For each job $j \in \mc J$, let $p'_j$ be its processing time scheduled on machines of $\Mnt$ in $\solopt$, and $p''_j = p_j - p'_j$.
For each class $i$, let $P'(C_i) = \sum_{j \in C_i} p'_j$.
We can make the following statement about the number of setup times $s_i$ per class $i$ on $\Mnt$ in $\solopt$.
\begin{lemma}
	\label{lem:nine:number_of_setups}
	The number $\lambda'_i$  of setup times $s_i$ on $\Mnt$ in $\solopt$ is at least $\ceil*{\sfrac{P'(C_i)}{(1/3-s_i)}}$.
\end{lemma}
\begin{proof}
	By definition of the machines in $\Mnt$, at most $1/3 - s_i$ processing time of job pieces of any class $i$ can be scheduled on each machine in $\solopt$.
	Thus, $P'(C_i)$ is processed on at least $\ceil*{\sfrac{P'(C_i)}{(1/3-s_i)}}$ different machines of $\Mnt$ in $\solopt$, requiring at least one setup time per machine.
\end{proof}
This statement also implies that there is at least one setup $s_i$ solely for each job $j$ of class $i$ with $s_j + p'_j > \sfrac13$ on $\Mnt$.
We start by removing all job pieces scheduled on machines of $\Mnt$ in $\solopt$, putting them into a set $\Snt$.
As all jobs of classes in \typeseven on machines in $\Mnt$ are scheduled in the intervals $[0,\sfrac13]$ and $[\sfrac23,1]$, there is at most $\sfrac23 - 2 s_i$ processing time of a single job $j$ of class $i$ scheduled on machines in $\Mnt$, and every job $j$ with $s_i + p'_j > \sfrac13$ is scheduled on at least two different machines, and has job pieces in both intervals $[0,\sfrac13]$ and $[\sfrac23,1]$ in $\solopt$.

\begin{enumerate}[]
\item \case[1]{$\exists j \in \Snt : 2s_i + p_j \geq \sfrac23 \wedge s_i + p'_j > \sfrac13$.}
\label{case:nine:g13_g23}

\begin{addmargin}[2em]{0em}
Note that, for each such $j$, $p''_j > 0$, and therefore there is a setup time $s_i$ of its class $i$ already included in $\Inice$.
Let $J \subseteq \Snt$ be the set of all such jobs.
For each class $i$, let $\Pover(C_i) = \sum_{j \in C_i \setminus J} p'_j - (\ceil*{\sfrac13 - s_i} - 2 \abs{C_i \cap J})$, \ie the total processing time of job pieces of class $i$ in $\Snt$ that have no setup time if all jobs of $C_i \cap J$ are placed with exactly two setup times $s_i$.
By \autoref{lem:nine:number_of_setups}, at least $\Pover(C_i)$ processing time of job pieces of class $i$ then must be placed together with jobs in $C_i \cap J$ on machines $\Mnt$ in $\solopt$.

We start by ordering the machines $q$ in $\Mnt$ by non-increasing $F_q = 1-L(\inineten)$, where $\inineten$ is the class of $\typenine \cup \typeten$ scheduled on a machine in $\solopt$.
Let $\Mnt'$ be the set of the first $2 \abs{J}$ machines by the ordering, and $F = \sum_{q \in \Mnt'} F_q$.
Then, $\sum_{j \in J} (2s_j + p'_j)  +  \sum_{i \in \iv c} \Pover(C_i) \leq F$:
Let $q,q'$ be the last two machines added to $\Mnt'$.
Then, by the ordering of machines, the largest non-overlapping interval not filled with classes of $\typenine \cup \typeten$ over all machines in $\Mnt \setminus \Mnt'$ has size at most $F_q + F_{q'}$.
Let $F' = F - \frac{\abs{\Mnt'}}{2} (F_q + F_{q'})$ be the total free space of machines $\Mnt'$ outside the interval $F_q + F_{q'}$.
Then, at most $2 \abs{J} (F_q + F_{q'})$ total load of $\sum_{j \in J} (2s_j + p'_j)  +  \sum_{i \in \iv c} \Pover(C_i)$ can be placed in the interval $(F_q + F_{q'})$ over all machines in $\Mnt$, and the rest must be placed in $F'$, as otherwise, either some job $j \in J$ would be parallelized in $\solopt$, or there would be more than $\lambda'_i$ setup times for some class $i$ in $\solopt$.
As $F = F' + \sfrac{\abs{\Mnt'}}{2} (F_q + F_{q'})$,  $\sum_{j \in J} (2s_j + p'_j)  +  \sum_{i \in \iv c} \Pover(C_i) \leq F$ follows.

We order these jobs in $J$ by non-decreasing setup time $s_i$ of their respective classes $i$.
While for the first $j \in J$, $F_q + F_{q'} \geq s_i + \sfrac13$ for the first two machines $q,q' \in \Mnt'$, we know that, using the extra $\sfrac13$ space given per machine by only approximating the optimal solution, the total free space on these two machines is at least $\sfrac23 + F_q + F_{q'} \geq \sfrac23 + s_i + \sfrac13 = s_i + 1$.
As it holds for all $j$ that $p_j \leq 1 - s_i$ by definition of $\solopt$, we can schedule the whole job $j$ on $q$ and $q'$ together with $\inineten$ and $\inineten'$, the classes of $\typenine \cup \typeten$ scheduled on $q$ and $q'$ in $\solopt$.
To do this without parallelization of $j$, we first schedule $\inineten$ on $q$ such that it finishes at exactly $\sfrac43$, $\inineten'$ on $q'$ such that it starts at $0$, proceeding with $j$ by scheduling $s_i$ as well as at most $\sfrac43 - L(\inineten) - s_i$ processing time of $j$ on $q$ starting at $0$, and $s_i$ as well as $p_j - (\sfrac43 - L(\inineten) - s_i)$ processing time of $j$ on $q'$, starting at $L(\inineten')$.
Since $2s_i + p_j \geq \sfrac23$, the total load of setup times and job pieces on $q$ and $q'$ is at least $2$ after this process.

Afterwards, we add $j$ to a set $\Jone$, and $q,q'$ to a set $\Mlarge$.
Let $\Jtwo$ be the set of jobs not placed by this process.
It now holds, by the ordering of $J$ and $\Mnt'$, that $F_q + F_{q'} < s_j + \sfrac13$, for all $j \in J$ and $q,q' \in \Mnt' \setminus \Mlarge$.
As all jobs of $\Jone$ have been scheduled on exactly two machines of $\Mnt'$, $\Mnt' \setminus \Mlarge = 2 \abs{\Jtwo}$.
Thus, $\sum_{j \in \Jtwo} s_j + \sfrac13 > \sum_{q \in \Mnt' \setminus \Mlarge} F_q$.

Let $\Pext = \sum_{j \in \Jone} p''_j$.
As now $L(q) > 1$ for each $q \in \Mlarge$,

\begin{equation*}
\begin{aligned}
&\Pext + \sum_{j \in \Jone} (2s_j + p'_j) + \sum_{j \in \Jtwo} (s_j + \sfrac13)\\
&> \sum_{q \in \Mlarge} F_q + \sum_{q \in \Mnt' \setminus \Mlarge} F_q = F \\
&\geq  \sum_{j \in J} (2s_j + p'_j)  +  \sum_{i \in \iv c} \Pover(C_i)\\
\Leftrightarrow &\Pext >  \sum_{j \in \Jtwo} (s_j + p'_j - \sfrac13)  +  \sum_{i \in \iv c} \Pover(C_i)
\end{aligned}
\end{equation*}

As exactly $\Pext$ load is removed from $\Snice$ when scheduling the jobs from $\Jone$, we can safely move $s_j + p'_j - \sfrac13$ processing time of each $j \in \Jtwo$, as well as $\Pover(C_i)$ processing time of some jobs $j \in \Snt$ of each class $i$, from $\Snt$ to $\Snice$ without violating $\Lnice(\Inice) \leq \abs{\Mnice}$.
Then, the remaining processing time of each $j \in \Jtwo$ is at most $p'_j - (s_j + p'_j - \sfrac13) \leq \sfrac13 - s_j$.
Now, all jobs $j$ with previously $s_i + p'_j > \sfrac13 \wedge 2s_i + p_j \geq \sfrac23$ have either been fully placed on machines in $\Mlarge$ (the jobs in $\Jone$), or it now holds for such jobs that $s_j + p'_j \leq \sfrac13$ (the jobs in $\Jtwo$), and there are thus no more jobs with $s_j + p'_j > \sfrac13$ to consider.
Additionally, for all $i \in \iv c$, by this we ensure $\Pover(C_i) \leq 0$, \ie there are at least $\ceil*{\sum_{j \in C_i} p'_j / (\sfrac13-s_i)}$ setup times left in $\Snt$ for each class $i$.
\end{addmargin}

\item \case[2]{$\exists j \in \Snt : 2s_i + p_j > \sfrac23 \wedge s_i + p'_j \leq \sfrac13 \wedge \Mnt \neq \emptyset$.}
\label{case:nine:leq13_g13}

\begin{addmargin}[2em]{0em}
We remove machine $q$ from $\Mnt$ with largest $F_q = 1-L(\inineten)$, where $\inineten$ is the class of $\typenine \cup \typeten$ scheduled on a machine in $\solopt$, schedule the class $\inineten$ on $q$ starting at $\sfrac13$, and schedule $s_i$ as well as $\sfrac13 - s_i$ processing time of $j$ directly below on $q$, starting at $0$.
To ensure $\ceil*{\sum_{j \in C_i} p'_j / (\sfrac13-s_i)}$ setup times are left for $i$ in $\Snt$ afterwards, we remove $\sfrac13 - s_i - p'_j$ processing time of some jobs $j'$, $j' \in C_i$, from $\Snt$ if possible, and add it to $\Snice$.
As for such $j$, $p''_j > 0$, there is a setup time $s_i$ already included in $\Snice$, and we have just scheduled $\sfrac13 - s_i - p'_j$ processing time from $\Snice$ on $q$, this preserves $\Lnice(\Inice) \leq \abs{\Mnice}$ as well. 
\end{addmargin}
\end{enumerate}
Afterwards, it holds for all $j \in \Snt$ that $2s_i + p_j \leq \sfrac23$.
On all machines $1, \dots, \abs{\Mnt} \in \Mnt$, we schedule all remaining classes $\inineten \in \typenine \cup \typeten$ alternately starting at $0$ or finishing at $\sfrac43$, such that on machine $\abs{\Mnt}$, $\inineten$ finishes at $\sfrac43$.
This schedules all remaining classes in $\typenine \cup \typeten$ by definition of $\Mnt$.
We define a Wrap Template $\omega = (\omega_1, \dots, \omega_{\abs{\Mnt}})$, with $a_r = 0$ and $b_r = \sfrac43 - L(\inineten)$ if $\inineten$ finishes at $\sfrac43$ on machine $r$, or $a_r = \sfrac43 - L(\inineten)$ and $b_r = \sfrac43$ if $\inineten$ starts at $0$ on machine $r$.
Let $F  = \sum_{r = 1}^{\abs{\omega}} b_r - a_r - \sfrac13$ the total space on machines $1, \dots, \abs{\omega}$ available in $\solopt$ besides the classes of $\typenine \cup \typeten$.
Let $Q$ be an empty Wrap Sequence.
We first execute \autoref{algo:9_structural_wrapseq} to fill $Q$.
\begin{algorithm}[h!]
\caption{}\label{algo:9_structural_wrapseq}
\begin{algorithmic}[1]
\While{$\Snt \neq \emptyset$ and $F > 0$}
\State Let $i$ be a class with $s_i \in \Snt$.
\State Move $s_i$ from $\Snt$ to $Q$ and set $F := F - s_i$.
\While{$F > 0$ and $\exists j : j \in C_i \wedge p'_j > 0$}
\State Remove $p'_j$ from $\Snt$, $p''_j$ from $\Snice$.
\If{$F > p_j$}
\State Add $j$ to $Q$ and set $F = F - p_j$.
\Else
\State Split $p_j$ into two parts $j^1$ and $j^2$ with $p_{j^1} = F$ and $p_{j^2} = p_j - F$.
\State Add $j^1$ to $Q$, $j^2$ to $\Snice$, and set $F := 0$.
\EndIf
\EndWhile
\EndWhile
\end{algorithmic}
\end{algorithm}
We then execute \wrapdyn{}($Q,\omega$). \autoref{prop:wrapdyn:space} is fulfilled: $L(Q) \leq S(\omega) - \sum_{r=2}^{\abs{\omega}} s_r$ is true as $\sum_{r=2}^{\abs{\omega}} s_r \leq (\abs{\omega} - 1) \sfrac13$ by definition of classes in $\Snt$, $L(Q) = F$ by definition of the procedure above, and $S(\omega) = F + (\abs{\omega}-1) \sfrac13$, as well as $b_r - a_r \geq \sfrac13 \geq \max_{r \in \abs{\omega}} s_r$ by definition.
The schedule produced by \wrapdyn{}($Q,\omega$) is also never parallelized, as only jobs $j$ of classes $i$ with $2s_i + p_j \leq \sfrac23$ are placed, and, by our definition of $\omega$, no matter where a job begins, the next $\sfrac23$ space in $\omega$ does not overlap.
Note that, for every class $i$ with $s_i + P(C_i) > \sfrac13$ and $p'_j > 0$, there are at least two setup times scheduled in $\solopt$: Either two $s_i$ are needed to schedule $i$ fully on machines in $\Mnt$, or $i$ is only partly scheduled on machines in $\Mnt$, and thus, $s_i$ is already included also in $\Snice$.
As by definition of \wrapdyn, we only use one \emph{real} $s_i$ for each such class when wrapping it, as the total sum of all additional setup times is lesser or equal than the additional space, we can include one $s_i$ for each such $i$ in $\Snice$ without increasing the total load to be scheduled.

If, after this process, $\Snt = \emptyset$, we are done: All jobs previously (partly) scheduled on machines in $\Mnt$ are either fully scheduled on these machines, or below $\sfrac13$ if they had less than $\sfrac13$ placed on machines in $\Mnt$ in $\solopt$ (\caseref{case:nine:leq13_g13}{2}).
The resulting schedule is feasible and has makespan at most $\sfrac43$ by construction.
Simultaneously, we have ensured at every step that $\Lnice(\Inice) \leq \abs{\Mnice}$, and we are thus able to compute a $\sfrac43$-approximation for $\Inice$ in time $\Oh(n)$ by \autoref{preemptive:simple}.
As all jobs of classes in \typeseven that may have job pieces in both $\Snt$ and $\Snice$ are jobs considered in \caseref{case:nine:leq13_g13}{2}, and the pieces in $\Snt$ are feasibly placed below, the pieces in $\Snice$ feasibly placed above $\sfrac13$, the resulting schedule has no parallelization, and is thus a feasible $\sfrac43$-approximation.

Let us thus consider the case that $\Snt \neq \emptyset$ after the process.
We add $\Snt$ to $\Snice$, removing all duplicate setup times from $\Snice$.
Then there is at most one class $e$ with a higher number of setup times in our current solution compared to the original solution $\solopt$:
For all classes $i$ that are fully scheduled on machines in $\Mnt$, we use at most $\lambda'_i$ setup times, by ensuring in \caseref{case:nine:g13_g23}{1} and \caseref{case:nine:leq13_g13}{2} that there always are enough setup times left afterwards to schedule the remaining job pieces of each class, and only including jobs of classes $i$ in $Q$ which have at least one setup time $s_i$ still left in $\Snt$ after these cases.
All classes $i$ that already had some job pieces in $\Snice$ before $\Snt$ was added to $\Snice$, or that are now fully included in $\Snice$, have exactly one setup time in $\Snice$.
Only at most a class $e$, the class which jobs are added last to $Q$ in \autoref{algo:9_structural_wrapseq}, may now have an additional setup time in $\Snice$ if $e$ was partly added to $Q$ and partly left in $\Snt$.
However, this setup time is only additional iff $s_e + P(C_e) > \sfrac13$, as argued above.
Thus, if $e$ now actually has more setup times than in $\solopt$, we can conclude that $s_e + P(C_e) \leq \sfrac13$ by \autoref{lem:nine:number_of_setups}.
As there is at least $\sfrac13$ free space on machine $\omega_1$ by definition of $\omega$, scheduling $e$ fully on this machine and removing $s_e$ from $\Snice$ implies that the total number of setup times of each class in our current solution is lesser or equal than this number of each class in $\solopt$.
This implies that the total load of $\typeseven$ classes in our current solution is lesser or equal than the total load of these classes in $\solopt$.
Since $L(Q) = F$, it then must hold that $\Lnice(\Snice) \leq \abs{\Mnice}$, as otherwise, the total load of all setup times and job pieces placed in $\solopt$ is greater than $m$, a contradiction to the optimality of the solution.
Now, we only have to make sure that can not occur any parallelization for all classes.
This is already fulfilled for all classes but $e$: Besides at most one job $j \in C_e$, there are only full classes and full jobs on machines in $\Mnt$; only $j$ can have a job piece in both $Q$ and $\Snice$ which can possible overlap.
This can only happen if $s_e + P(C_e) > \sfrac13$, as otherwise, we have already included $e$ fully in $Q$.
Let $p'_j$ be the processing time of all job pieces of $j$ in $Q$, $p''_j$ the processing time of all job pieces of $j$ in $\Snice$.
We slightly modify $\omega$ by setting $a_1 = 0$, \ie increasing the total space of $\omega$ by $\sfrac13$.
As $2s_i + p_j \leq \sfrac23$, either $p'_j \leq \sfrac13$ or $p''_j \leq \sfrac13$ must hold.
If $p''_j \leq \sfrac13$, we can then simply add $j$ fully to $Q$ without violating any of the Batch Wrapping properties.
If instead $p'_j \leq \sfrac13$, we modify $\omega$ again by setting $b_{\abs{\omega}} = \sfrac13$.
This still guarantees that $S(\omega) \geq F$, but now also makes sure that all of $p'_j$ is scheduled in the interval $[0,\sfrac13]$ on machine ${\abs{\omega}}$; as \autoref{preemptive:simple} places all jobs of \typeseven classes above $\sfrac13$, this guarantees that no parallelization occurs.
We are thus able to compute a feasible $\sfrac43$-approximation for $\Inice$ via  \autoref{preemptive:simple}, and thus, a $\sfrac43$-approximation for the whole instance.
The resulting schedule, in either case, has the following properties:
\begin{itemize}
\item all classes of $\{\typeone, \dots, \typesix\}$ and some parts of classes of $\typeseven$ are part of a nice instance
\item all classes of \typeeight are scheduled together with classes of \typenine  with no other setup times or job pieces on one machine each
\item all other \typenine classes and all \typeten classes are scheduled on a machine each
\item all remaining jobs $j$ of \typeseven classes are either fully scheduled machines containing \typenine or \typeten classes, respectively, or scheduled partly on one machine containing a \typenine or \typeten class below $\sfrac13 T$ iff $2s_i + p_j > \sfrac23 T$
\end{itemize}
	
\section{The Approximation Algorithm}
\label{sec:approxalgo}

In this section, we formally describe the $(\sfrac43+\eps)$-approximation algorithm.
Due to preemption, the optimal makespan $\OPT$ is not necessarily be integral.
One can thus not simply guess all integral values between some lower and upper bound on $\OPT$ via binary search.
Instead, we allow a multiplicative error of $\eps$, for any $\eps > 0$, on $\OPT$, \ie guess some value $\OPT \leq T \leq (1 + \eps) \OPT$.
Any $\sfrac43 T$-approximate solution then is a $(\sfrac43 + \eps)$-approximate solution w.r.t the optimal makespan $\OPT$.
We do this by computing a $2$-approximation \cite{DJ19} in time $\Oh(n)$ with some makespan $\OPT \leq T' \leq 2 \OPT$.
We set a lower bound $\ell = T'$, and an upper bound $u = 2 T'$.
Then, using Binary Search, we need at most $\Oh(\log(1/\eps)$ iterations to find some lower bound $\ell'$ and upper bound $u'$ with $(1 + \eps) \ell' \geq u' \geq \ell'$, such that no solution with makespan $\ell'$ exists, while we can compute a solution with makespan at most $\sfrac43 u'$, implying that $u' \leq (1 + \eps) \OPT$.

We now give a description of the $\sfrac43T$-approximation algorithm for general instances, for some fixed makespan guess $T$, which we run for each guess of the binary search.
If there exists an optimal solution with makespan $T$, our algorithm will compute a $\sfrac43T$-approximate solution, which we call $\sigma$; otherwise, it will allow us to conclude that no such solution exists.
Again, we normalize the instance $I$ by dividing all input parameters, as well as $T$ itself, by $T$.
Given an instance $I$ of \PCmaxpm, we structure $I$ into types $\{\typeone, \dots, \typeten\}$ as defined in \autoref{tab:preemptive:class_partition}, in time $\Oh(c)$.
Let $\solstr$ be the $\sfrac43$-approximate solution with properties as described in \autoref{sec:structure}, following either \autoref{sec:eight_rem} or \autoref{sec:nine_rem}.
If $\solstr$ does not exist, there also does not exist a solution with makespan at most $T$, as shown in \autoref{sec:structure}.
We will compute a solution $\sigma$ with these properties, while minimizing the total load of setup times over all such solutions, and thus the total load of setup times and job pieces to be distributed on the $m$ machines.
This allows us to conclude that, if we are not able to compute $\sigma$, $\solstr$ does not exist for $I$, implying that there exists no solution with makespan $T$ for $I$ at all.

We begin the construction of $\sigma$ by scheduling pairs of classes $\ieight \in \typeeight$ and $\inine \in \typenine$ on a single machine starting at $\sfrac13$, respectively, until either $\abs{\typeeight} = 0$ or $\abs{\typenine} = 0$, in time $\Oh(c)$.  
If there are not enough machines to do this, we can already conclude that there exists no schedule with makespan $T$.
This also tells us which case we are in, \ie $\abs{\typenine} = 0$ after this process, $\solstr$ must have the properties as described in \autoref{sec:eight_rem}; if $\abs{\typeeight} = 0$ after this process, $\solstr$ must have the properties as described in \autoref{sec:nine_rem}.

\subsection{Classes of Type Eight Remain}
\label{sec:algo:eight_rem}
Then \textbf{$\abs{\typenine} = 0$.} We continue by constructing a bipartite graph, with a vertex for each class $\ieight \in \typeeight$ and each class $\iten \in \typeten$, and an edge between two vertices iff one vertex represents a class of \typeeight, the other a class of \typeten, and the combined load both classes is lesser or equal than $\sfrac43$.
We construct this graph and compute a maximal matching for it in time at most $\Oh(c^2)$~\cite{F56}.
We schedule all classes of $\ieight \in \typeeight$ and $\iten \in \typeten$ corresponding to the vertices of an edge included in the maximal matching together on a single machine.

All pairs $\ieight,\ieight' \in \typeeight$ that are not yet scheduled, we schedule together on a single machine starting at $\sfrac13$, leaving at most one class $\ieight^* \in \typeeight$ unscheduled.
All remaining classes $\iten \in \typeten$ are scheduled alone a machine, each with starting time $\sfrac13$.
We call the set of machines with pairs of \typeeight classes and single \typeten classes scheduled in $\sigma$ on them $\Met$.
If this is not possible, $\solstr$ does not exist.
This leaves no class in $\typeten$, and at most one class $\ieight \in \typeeight$, unscheduled.
The remaining machines with no setup times or job pieces scheduled on them yet we call $\Mrem$.

Let $\Sseven$ be the set of classes $i \in \typeseven$ of $I$.
We know that, if $\solstr$ exists, there exists a split of $\Sseven$ into two sets $S_1,S_2$ of (partial) classes, with a setup time $s_i$ in each set where $i$ is (partly) included, such that all setup times and job pieces of $S_1$ are scheduled in intervals $[0,\sfrac13]$ of machines in $\Met$, and all setup times and job pieces of $S_2$ are either part of a nice instance, or scheduled in pairs on machines.

In fact, $s_i \leq \sfrac16$ for all setup times that are part of $S_1$, and $S_1$ can be further divided into sub-sets $S_1^{> \sfrac19}$, containing all setup times and job pieces of classes $i$ with $s_i > \sfrac19$, and $S_1^{\leq \sfrac19}$ containing $S_1 \setminus S_1^{> \sfrac19}$.
According to $\solstr$, there exists some $0 \leq r' \leq \abs{\Met}$, such that some classes $i  \in S_1^{> \sfrac19}$ with $\sfrac16 < s_i + P(C_i) \leq \sfrac13$ are fully scheduled on a number of machines $r_1' \leq r'$, some classes $i  \in S_1^{\leq \sfrac19}$ with $\sfrac29 < s_i + P(C_i) \leq \sfrac13$ are fully scheduled on a number of machines $r_2' \leq \abs{\Met}-r'$, and it holds for the sum of loads $L_1$ of all remaining partial classes $i \in S_1^{> \sfrac19}$ and $L_2$ of all remaining partial classes $i' \in S_1^{\leq \sfrac19}$ that

$$L_1 + L_2 \leq \sfrac16 (r'-r_1) + \sfrac29 (\abs{\Met} - r_2 - r') \enspace .$$

$S_2$ can be further divided into two sets $S'_2,S''_2$ such that $S'_2$ is part of a nice sub-instance in $\solstr$, and $S''_2$ contains a number of classes $i$ with $\sfrac12 < s_i + P(C_i) \leq \sfrac59$ where an even number of these classes is scheduled in pairs on machines in $\Mrem$, and one such class is scheduled together with $\ieight^*$ on a machines in $\Mrem$ if $S''_2$ is odd.

We know aim to compute a split of $\Sseven$ into two sets $R_1,R_2$ akin to $S_1,S_2$, such that the load of $R_1$, \ie the load placed on machines $\Met$, is maximized over all such splits, while ensuring that the total load $L(R_1 + R_2)$ to be distributed stays small enough such that $R_2$ can still be feasibly placed on machines in $\Mrem$.
As the total load to be distributed depends only on the number of setup times used in a solution, this essentially boils down to minimizing the sum of all setup times in $R_1 \cup R_2$.

For each class $i \in \iv c$, we compute the \enquote{obligatory} minimum load of each job $j$ that must be scheduled as part of $S_2$: For each $j$ of class $i$, at most $\sfrac13-s_i$ processing time can be scheduled on machines in $\Met$ by the properties of $\solstr$.
For each job $j$, this can be expressed as $p'_j = \max(p_j - \sfrac13 + s_i),0)$, and as $P'(C_i) = P(C_i) - \sum_{j \in C_i} p'_j$ for each class $i$.
We compute three ordered partial copies of $\Sseven$ in time $\Oh(c \log(c))$:
One where all classes $i$ with $s_i \leq \sfrac16$ and $P'(C_i) > 0$ are ordered non-decreasing by $\sfrac{s_i}{P'(C_i)}$,
one where all classes $i$ with $\sfrac16 < s_i + P(C_i) \leq \sfrac13$ are ordered non-increasing by $s_i + P(C_i)$,
and one where all classes $i$ with $s_i \leq \sfrac19$ as well as $\sfrac29 < s_i + P(C_i) \leq \sfrac13$ are ordered non-increasing by $s_i + P(C_i)$.

We guess $r'$ in at most $c$ guesses.
We set $r_1 = r_2 = 0$.
While there still exists a class $i$ with $s_i > \sfrac19$ and $\sfrac16 < s_i + P(C_i) \leq \sfrac13$, and $r_1 < r'$, we remove the largest such $i$ from $\Sseven$, schedule in the interval $[0,\sfrac13]$ on some machine $q \in \Met$, and increase $r_1$ by $1$.
Similarly, while there still exists a class $i$ with $s_i \leq \sfrac19$ and $\sfrac29 < s_i + P(C_i) \leq \sfrac13$, and $r_2 < \abs{\Met} - r'$, we remove the largest such $i$ from $\Sseven$, schedule in the interval $[0,\sfrac13]$ on some machine $q \in \Met$, and increase $r_2$ by $1$.
Afterwards, we set $\omega = \abs{\Met} - r_2 - r_1$, \ie the number of machines in $\Met$ where the interval $[0,\sfrac13]$ is still empty, and set $r' = r' - r_1$.
Let $F_1 : = \sfrac16 r'$ and $F_2 := \sfrac29 (\abs{\omega} - r')$.

Now, consider \autoref{algo:classes_below}.
\begin{algorithm}[!ht]
\caption{$(Q_k,F,\ell,u)$}\label{algo:classes_below}
\begin{algorithmic}[1]
\While{$\exists i : \ell < s_i \leq u \wedge P'(C_i) = 0$ and $F > 0$}
\State Add $s_i$ to $Q_k$.
\If{$P(C_i) \leq F$}
\State Add $C_i$ to $Q_k$.
\State Set $F = F - s_i - P(C_i)$.
\Else
\State Add a set of job pieces $J' \in C_i$ to $Q_k$, with $P(J') = F$.
\State Set $F = 0$.
\EndIf
\EndWhile
\While{$\exists i : \ell < s_i \leq u \wedge P'(C_i) > 0$ and $F > 0$}
\State Let $i$ be the class with smallest $\sfrac{s_i}{(P(C_i)-P'(C_i))}$ over all such classes.
\State Add $s_i$ to $Q_k$.
\If{$P(C_i)-P'(C_i) \leq F$}
\State For each job $j \in C_i$, add a job piece $j'$ with $p_{j'} = p'_j$ to $Q_k$.
\State Set $F = F - s_i - (P(C_i)-P'(C_i))$.
\Else
\State For each job $j \in C_i$, add a job piece $j'$ with $p_{j'} \leq p'_j$ to a set $J'$, until $P(J') = F$.
\State Add $J'$ to $Q_k$.
\State Set $F = 0$.
\EndIf
\EndWhile
\end{algorithmic}
\end{algorithm}
We execute \autoref{algo:classes_below} first for the parameters $(Q_2,F_2,0,\sfrac19)$, then for $(Q_1,F_1,\sfrac19,\sfrac16)$, and, if afterwards $L(Q_1) < F_1$, again for $(Q_1,F_1-L(Q_1),0,\sfrac19)$, in time $\Oh(n)$ each.
$Q_2$ contains at most all non-obligatory load of classes $i$ with $s_i \leq \sfrac19$, $Q_1$ at most all non-obligatory load of classes $i$ with $s_i \leq \sfrac16$.
Let $s_{e_1}$ and $s_{e_2}$ be the setup times of the last classes $C_{e_1}$ and $C_{e_2}$ added to $Q_1$ and $Q_2$, respectively.
We redefine $Q_1, Q_2$ as Wrap Sequences from their respective sets, starting with $s_{e_1}$ and $s_{e_2}$, respectively, with the rest of the (partial) classes in arbitrary order.
We create two Wrap Templates $\omega^1,\omega^2$, with $\abs{\omega^1} = r'$, $\abs{\omega^2} = \abs{\omega} - r'$, and $\omega^i_r = (r, 0, \sfrac13)$ for $i \in \{0,1\}$ and $1 \leq r \leq \abs{\omega^i}$.
We execute both \wrapdyn($Q_1,\omega^1$) and \wrapdyn($Q_2,\omega^2$) in time at most $\Oh(c)$.
By definition of \autoref{algo:classes_below} and the fact that $s_{e_1} \leq \sfrac16$, we know that
\begin{equation*}
\begin{aligned}
L(Q_1) &\leq \sfrac16 \abs{\omega^1} + s_{e_1} = \sfrac13 \abs{\omega^1} - \sfrac16 \abs{\omega^1} + s_{e_1} \\
&\leq \sfrac13 \abs{\omega^1} - \sfrac16 (\abs{\omega^1}-1) \leq S(\omega^1) - \sum_{r=2}^{\abs{\omega^1}} s_r \enspace .
\end{aligned}
\end{equation*}
The same is true for $L(Q_2)$, as $s_{e_2} \leq \sfrac19$ and thus
\begin{align*}
L(Q_2) &\leq \sfrac29 \abs{\omega^2} + s_{e_2} = \sfrac13 \abs{\omega^2} - \sfrac19 \abs{\omega^2} + s_{e_2} \\
&\leq \sfrac13 \abs{\omega^2} - \sfrac19 (\abs{\omega^2}-1) \leq S(\omega^2) - \sum_{r=2}^{\abs{\omega^2}} s_r \enspace .
\end{align*}
This implies that \autoref{prop:wrapdyn:space} is fulfilled for both $(Q_1,\omega^1)$ and $(Q_2,\omega^2)$.
\autoref{prop:wrapdyn:para} is fulfilled solely by definition of \autoref{algo:classes_below}, as we only add non-obligatory load of jobs, \ie at most $\sfrac13 - s_i$ processing time per job, to $Q_1$ and $Q_2$ respectively.

Let $R_1$ be the set of all setup times and job pieces of classes in $\typeseven$ now placed on machines $\Met$, and $R_2$ the set of all job pieces not in $R_1$, as well as one setup time per class of the job pieces in $R_2$.
We now aim to prove $L(R_2) \leq L(S_2)$, which will allow us to place $R_2$ feasibly on $\Mrem$.
In the following, we first assume that $L(Q_1) = F_1$ and $L(Q_2) = F_2$.
If $r'$ was guessed correctly, we now have $L(R_1) \geq L(S_1) + s_{e_1} + s_{e_2}$: 
By first placing always the largest $i$ with either $s_i > \sfrac19$ and $\sfrac16 < s_i + P(C_i) \leq \sfrac13$ or  $s_i \leq \sfrac19$ and $\sfrac29 < s_i + P(C_i) \leq \sfrac13$, we ensure that on both $r_1'$ and $r'_2$ machines in $\sigma$, there is at least as much load of $\Sseven$ placed as on $r_1'$ and $r'_2$ machines in $\solopt$.
On all $r' + (r_1 - r_1')$ machines, which have at most $F_1 + (r_1 - r_1') \sfrac16$ load of $\Sseven$ in $\solopt$, we place at least $F_1 + s_{e_1} + (r_1 - r_1') \sfrac16$ load of $\Sseven$ per machine, $F_1 + s_{e_1}$ by the execution of \autoref{algo:classes_below}, and at least $(r_1 - r_1') \sfrac16$ by placing $i$ with $\sfrac16 < s_i + P(C_i) \leq \sfrac13$ on machines $(r_1 - r_1')$.
Similarly, for all $r' + (r_2 - r_2')$ machines, which have at most $F_2 + (r_2 - r_2') \sfrac29$ load of $\Sseven$ in $\solopt$, we place at least $F_2 + s_{e_2} + (r_2 - r_2') \sfrac29$ load of $\Sseven$ per machine, $F_2 + s_{e_2}$ by the execution of \autoref{algo:classes_below}, and at least $(r_2 - r_2') \sfrac29$ by placing $i$ with $\sfrac29 < s_i + P(C_i) \leq \sfrac13$ on machines $(r_2 - r_2')$.

With this in hand, we can show that $L(R_2) \leq L(S_2)$: If there are no parts of classes $i$ with $P'(C_i) > 0$ in $R_1$, then the only setup times to appear in $R_1 \cup R_2$ more than once are $s_{e_1}$ and $s_{e_2}$.
Since we know that $L(R_1) \geq L(S_1) + s_{e_1} + s_{e_2}$ by definition of our procedure, this implies $L(R_2) \leq L(S_2)$.
Otherwise, we first add all possible classes with $P'(C_i) = 0$ to $R_1$, and afterwards, minimize the total sum of setup times appearing both in $R_1$ and $R_2$ by always adding all non-obligatory load of the class $i$ with current smallest $\sfrac{s_i}{(P(C_i)-P'(C_i))}$ to $R_1$, until $L(R_1) = F$.
Moreover, for all $0 \leq L \leq L(R_1)$, there exists some subset $R'_1 \subseteq R_1$ with $L (R'_1) = L$ that has the minimum total sum of setup times over all subsets with load $L$ and properties of $S_1$, by removing setup times and job pieces with a total of $L(R_1) - L$ load that were last added to $R_1$ by \autoref{algo:classes_below}, and thus also for some subset $R'_1 \subseteq R_1$ with $L(R'_1) = L(S_1)$.
Thus, $L(S_1 \cup S_2) \geq L(R'_1 \cup R_2)$.
As all setup times in job pieces in $R_1 \setminus R'_1$ can only reduce the load of $R_2$, and $L(R_1) = F_1 + F_2 + s_{e_1} + s_{e_2} \geq L(S_1)$, $L(R_2) \leq L(S_2)$ follows.

If instead both $L(Q_1) < F_1$ and $L(Q_2) < F_2$, all non-obligatory load of classes with $s_i \leq \sfrac16$ and $s_i \leq \sfrac19$ is contained in $R_1$, and, as $S_1^{> \sfrac19}$ and $S_1^{\leq \sfrac19}$ contain at most this load, respectively, both $S_1^{> \sfrac19} \subseteq R_1$ and $S_1^{\leq \sfrac19} \subseteq R_1$, immediately implying $L(R_2) \leq L(S_2)$.
If either $L(Q_1) < F_1$ or $L(Q_2) < F_2$, the proof of $L(R_2) \leq L(S_2)$ is a special case of the previous steps for $L(Q_1) = F_1$ and $L(Q_2) = F_2$, with $s_{e_1} = 0$ or $s_{e_2} = 0$ and considering the fact that $S_1^{> \sfrac19} \subseteq R_1$ or $S_1^{\leq \sfrac19} \subseteq R_1$, respectively.

Now, let $R''_2 \subseteq R_2$ be the subset of classes $i$ that are fully included in $R_2$ with $\sfrac59 \geq s_i + P(C_i) > \sfrac12$, and $R'_2 = R_2 \setminus R''_2$.
We schedule an even number of classes $i \in R''_2$ in pairs on machines of $\Mrem$.
If a single class $i \in R''_2$ remains afterwards, we schedule it together with $\ieight^*$ on a single machine of $\Mrem$, if $\ieight^*$ exists and $L(\ieight^*) + L(i) > 1$, and otherwise add both $i$ and $\ieight^*$ to $R'_2$.
We then create a nice instance $\Inice = (\Snice,\Mnice)$, with $\Snice$ consisting of all classes of types $(\typeone, \dots, \typesix)$, $\ieight^*$ if it exists and is not yet scheduled, and $R'_2$, and with $\Mnice$ consisting of all machines of $\Mrem$ with no setup times or jobs scheduled on them yet.

If $\abs{\Mnice} > \mnice(\Inice)$, we can conclude $\solstr$ does not exist: In $\solstr$, all classes of types $(\typeone, \dots, \typesix)$ are all part of a nice sub-instance $\Inice' = (\Snice',\Mnice')$ with $\abs{\Mnice'} = \abs{\Mrem} - \ceil*{\abs{S''_2}/2}$.
$\Snice'$ also includes all setup times and job pieces of $S'_2$, but not $S''_2$, and $\ieight^*$ only if $S''_2$ is even.
Let $m' = \ceil*{\abs{S''_2}/2}$.
Then, for each of the $m'$ machines of $\Mrem$ that have a pair of classes in $S''_2$ scheduled on it in $\solstr$, there is a $\sfrac19$ free space on machines in $\Met$ in $\solstr$: In $\solstr$, each time two classes in $S''_2$ are combined, a machine is added to $\Met$, but only setup times and job pieces with load $\sfrac19$ are added to $S_1$, allowing for a placement of $\sfrac29$ load on this machine.
Thus, $L(R_1) \geq L(S_1) + m' \sfrac19$, and $L(R_2) - m' \sfrac19 \leq L(S_2)$, which in turn implies $\Lnice(\Inice) - m' \sfrac19 \leq \Lnice(\Inice')$.
As there is at most $\sfrac{10}{9}$ load on each of the $m'$ machines, the total load on machines $\Mrem$ in $\solopt$ is $\Lnice(\Inice') + \sfrac{10}{9} m' \geq \Lnice(\Inice) - m' \sfrac19  + \sfrac{10}{9} m' = \Lnice(\Inice) + m'$.
As we schedule at least $1$ load on all machines in $\Mrem \setminus \Mnice$,
in total, this implies $\Lnice(\Inice) \leq \abs{\Mnice}$, if $\solstr$ exists.
Therefore, in any case, \autoref{preemptive:simple},~(\ref{preemptive:simple:decision:T_check_true}) is fulfilled, and we can compute a $\sfrac43$-approximation for the nice sub-instance in time $\Oh(n)$.
As all other machines have load at most $\sfrac43$, and we have now scheduled all jobs in the instance, the resulting schedule is a $\sfrac43$-approximation of the whole instance.
The run time is dominated by the matching sub-routines, which take time at most $\Oh(c^2) \in \Oh(n^2)$ each, resulting in a run-time of at most $\Oh(n^2)$ for the whole algorithm for some fixed $T$, and a total run time of $\Oh(n^2 \log(1/\eps))$.

\subsection{Classes of Type Nine Remain}
\label{sec:algo:nine_rem}
Then \textbf{$\abs{\typeeight} = 0$.}
We know that in $\solstr$, each class $\inineten \in \typenine \cup \typeten$ is scheduled on a single machine, together with some jobs $j$ of classes of $\typeseven$, and the rest of the machines are part of a nice instance $\Inice' = (\Snice',\Mnice)$, if it exists.
We thus set $\Mnice = m - \abs{\typenine \cup \typeten}$, and create a set $\Mnt$ from the remaining machines, where we schedule all classes $\inineten \in \typenine \cup \typeten$ on a single machine, respectively.
We now want to build a nice instance $\Inice = (\Snice,\Mnice)$ for $I$, such that $\Lnice(\Snice) \leq \Lnice(\Snice')$; this immediately implies the computability of a $\sfrac43$-approximation for $\Inice$ by the computability of such a solution for $\Inice'$ for $\solstr$, if it exists.
Let $\Sseven$ be the set of classes $i \in \typeseven$ of $I$.

We also know that, for each $i \in \typeseven$ with a job $j \in C_i : 2 s_i + p_j > \sfrac23$, that $s_i \in \Snice'$.
We thus also add $s_i$ to $\Snice$ for each such $i$.
In $\solstr$, there is some number $x$ such that only the $x$ machines $q$ of $\Mnt$ with largest $F_q = 1-L(\inineten)$ have jobs $j$ with $2 s_j + p_j > \sfrac23$ scheduled on them, fully or partly, together with the classes $\inineten$.
We add all such jobs $j$ with $2 s_j + p_j > \sfrac23$ to a set $J$, order $J$ by non-decreasing $s_j$, order all $q \in \Mnt$ non-decreasing by $F_q$, each in time $\Oh(n \log(n))$, and finally guess this number $x$ (at most $n$ guesses).
Let $\Mnt'$ be the set of the first $x$ machines of $\Mnt$.
While for the first $j \in J$, $F_q + F_{q'} \geq s_j + \sfrac13$ for the first two machines $q,q' \in \Mnt'$, we schedule $\inineten$ on $q$ such that it finishes at exactly $\sfrac43$, $\inineten'$ on $q'$ such that it starts at $0$, $s_j$ as well as at most $\sfrac43 - L(\inineten) - s_j$ processing time of $j$ on $q$ starting at $0$, and $s_j$ as well as $p_j - (\sfrac43 - L(\inineten) - s_j)$ processing time of $j$ on $q'$, starting at $L(\inineten')$.
If, after this process, $\Mnt' \neq \emptyset$, we schedule a setup $s_j$, as well as $\sfrac13 - s_j$ processing time of the first $j \in J$ on the first machine $q \in \Mnt'$, starting at $0$, and $\inineten$ starting at $\sfrac13$.
We remove all such placed job pieces from $\Sseven$.

We set $\Mnt = \Mnt \setminus \Mnt'$, and proceed similar to $\solstr$:
On all machines $1, \dots, \abs{\Mnt} \in \Mnt$, we schedule all remaining classes $\inineten \in \typenine \cup \typeten$ alternately starting at $0$ or finishing at $\sfrac43$, such that on machine $\abs{\Mnt}$, $\inineten$ finishes at $\sfrac43$.
This schedules all remaining classes in $\typenine \cup \typeten$ by definition of $\Mnt$.
We define a Wrap Template $\omega = (\omega_1, \dots, \omega_{\abs{\Mnt}})$, with $a_r = 0$ and $b_r = \sfrac43 - L(\inineten)$ if $\inineten$ finishes at $\sfrac43$ on machine $r$, or $a_r = \sfrac43 - L(\inineten)$ and $b_r = \sfrac43$ if $\inineten$ starts at $0$ on machine $r$.
Let $F  = \sum_{r = 1}^{\abs{\omega}} b_r - a_r - \sfrac13$ the total space on machines $1, \dots, \abs{\omega}$ available in $\solopt$ besides the classes of $\typenine \cup \typeten$.
Let $Q$ be an empty Wrap Sequence.
We first execute \autoref{algo:9_algorithmic_wrapseq} to fill $Q$.
\begin{algorithm}[h!]
\caption{}\label{algo:9_algorithmic_wrapseq}
\begin{algorithmic}[1]
\While{$\exists i : s_i \notin Q \wedge s_i + P(C_i) \leq \sfrac13$ and $F > 0$}
\State \autoref{algo:9_algorithmic_addjobsofclass} ($i$).
\EndWhile
\While{$\exists i : s_i \notin \Snice \wedge s_i \notin Q \wedge s_i + P(C_i) > \sfrac13$ and $F > 0$}
\State Let $i$ be the class with smallest $s_i$ over all such classes.
\State \autoref{algo:9_algorithmic_addjobsofclass} ($i$).
\EndWhile
\While{$\exists i : s_i \in \Snice \wedge s_i \notin Q \wedge s_i + P(C_i) > \sfrac13$ and $F > 0$}
\State Let $i$ be the class with smallest $\sfrac{s_i}{(P(C_i)-\sum_{j \in J} p_j)}$ over all such classes.
\State \autoref{algo:9_algorithmic_addjobsofclass} ($i$).
\EndWhile
\end{algorithmic}
\end{algorithm}

\begin{algorithm}[h!]
\caption{($i$)}\label{algo:9_algorithmic_addjobsofclass}
\begin{algorithmic}[1]
\State Put $s_i$ into $Q$ and set $F := F - s_i$.
\While{$F > 0$ and $\exists j \in \Sseven \cap C_i$}
\State Remove $j$ from $\Sseven$.
\If{$F > p_j$}
\State Add $j$ to $Q$ and set $F = F - p_j$.
\Else
\State Split $p_j$ into two parts $j^1$ and $j^2$ with $p_{j^1} = F$ and $p_{j^2} = p_j - F$.
\State Add $j^1$ to $Q$, $j^2$ to $\Snice$, and set $F := 0$.
\EndIf
\EndWhile
\end{algorithmic}
\end{algorithm}
We then execute \wrapdyn{}($Q,\omega$). \autoref{prop:wrap:space} is fulfilled: $L(Q) \leq S(\omega) - \sum_{r=2}^{\abs{\omega}} s_r$ is true as $\sum_{r=2}^{\abs{\omega}} s_r \leq (\abs{\omega} - 1) \sfrac13$ by definition of classes in $\Sseven$, $L(Q) = F$ by definition of \autoref{algo:9_algorithmic_wrapseq} and \autoref{algo:9_algorithmic_addjobsofclass}, and $S(\omega) = F + (\abs{\omega}-1) \sfrac13$, as well as $b_r - a_r \geq \sfrac13 \geq \max_{r \in \abs{\omega}} s_r$ by definition.
The schedule produced by \wrapdyn{}($Q,\omega$) is also never parallelized, as only jobs $j$ of classes $i$ with $2s_i + p_j \leq \sfrac23$ are placed, and, by our definition of $\omega$, no matter where a job begins, the next $\sfrac23$ space in $\omega$ does not overlap.
We add all remaining jobs of $\Sseven$ to $\Snice$, and add a setup time for each class that has jobs but no setup time in $\Snice$, to $\Snice$.

Again, there is at most one job $j$ that is only partly included in $Q$, of a class $e$ that was considered last in \autoref{algo:9_algorithmic_wrapseq}, and there is at least $\sfrac13$ free space on machine $\omega_1$ by definition of $\omega$.
Let $p'_j$ be the processing time of all job pieces of $j$ in $Q$, $p''_j$ the processing time of all job pieces of $j$ in $\Snice$.
If $s_e + P(C_e) \leq \sfrac13$, we schedule $e$ fully on $\omega_1$.
Else, we slightly modify $\omega$ by setting $a_1 = 0$, \ie increasing the total space of $\omega$ by $\sfrac13$.
As $2s_i + p_j \leq \sfrac23$, either $p'_j \leq \sfrac13$ or $p''_j \leq \sfrac13$ must hold.
If $p''_j \leq \sfrac13$, we can then simply add $j$ fully to $Q$ without violating any of the Batch Wrapping properties.
If instead $p'_j \leq \sfrac13$, we modify $\omega$ again by setting $b_{\abs{\omega}} = \sfrac13$.
This still guarantees that $S(\omega) \geq F$, but now also makes sure that all of $p'_j$ is scheduled in the interval $[0,\sfrac13]$ on machine ${\abs{\omega}}$; as \autoref{preemptive:simple} places all jobs of \typeseven classes above $\sfrac13$, this guarantees that no parallelization occurs.

Now, it is only left to show that if $\solstr$ exists, indeed $\Lnice(\Inice) \leq \abs{\Mnice}$ holds.
As all machines of $\Mnt$ have load at least $1$ in our solution, $\Lnice(\Inice) > \abs{\Mnice}$ only if the total load of setup times in our solution is greater than in $\solstr$ and $L(Q) = F$.
When filling machines $\Mnt'$, there is exactly one setup time of some job $j \in J$ per machine of $\Mnt'$ in $\solstr$.
As we always schedule jobs of $J$ with smallest overall setup time first, we minimize the total load of setup times when scheduling jobs of $J$.

Let $Q'$ be the Wrap Sequence of $\solstr$ for Batch-Wrapping on machines $\Mnt \setminus \Mnt'$, and $\Inice' = (\Snice',\Mnice)$ the nice instance of $\solstr$.
When adding classes to $Q$ to be Batch-Wrapped on machines $\Mnt \setminus \Mnt'$, we first add all classes $i$ with $s_i + P(C_i) \leq \sfrac13$, then all classes $i'$ that have no job $j \in J$ in non-decreasing order of their setup, and finally, classes $i''$ that do have at least one job $j \in J$ in non-decreasing order of $\sfrac{s_i}{(P(C_i)-\sum_{j \in J} p_j)}$.
If $Q$ contains only classes $i$, then there is exactly one setup time per class in $Q \cup \Snice$, as $s_e + P(C_e) \leq \sfrac13$, which is the minimum number of setup times over all such solutions.
If $Q$ also contains only classes $i$ and $i'$, then $e = i'$ for some $i'$.
If $s_e$ both in $Q'$ and $\Snice'$, we are done.
If $s_e$ in $\Snice'$, but not in $Q'$, then either $L(Q) - s_e \geq L(Q')$, as in this case, $Q'$ contains only full classes and does not use the additional $\sfrac13$ space on machine $\omega_1$, or there is some $s_{e'}$ with $s_{e'} \in Q'$ and $s_{e'} \geq s_e$ by our ordering of $i'$.
In the former case, the total load of setup times in our solution is increased by $s_e$ compared to $\solstr$, but we schedule $s_e$ more load in $Q$ than in $Q'$, and thus, $\Lnice(\Inice) \leq \abs{\Mnice}$ if $\solstr$ exists.
In the latter case, the total load of setup times in our solution is smaller or equal than in $\solstr$, and $\Lnice(\Inice) \leq \abs{\Mnice}$ as well.
If $Q$ contains also classes $i''$, then $e = i''$ for some $i''$, and again, either $L(Q) - s_e \geq L(Q')$, or $Q'$ contains classes $i''$ as well.
By our ordering of $i''$ and prior scheduling of all classes $i$ and $i'$, the total load of setup times of classes $i''$ is minimized over all such solutions, and $\Lnice(\Inice) \leq \abs{\Mnice}$ if $\solstr$ exists. 

Thus, in any case, \autoref{preemptive:simple},~(\ref{preemptive:simple:decision:T_check_true}) is fulfilled, and we can compute a $\sfrac43$-approximation for the nice sub-instance in time $\Oh(n)$.
As all other machines have load at most $\sfrac43$, and we have now scheduled all jobs in the instance, the resulting schedule is a $\sfrac43$-approximation of the whole instance.
All orderings can be computed before guessing the size $x$ of $\Mnt'$ in time $\Oh(n \log(n))$.
The number of guesses on $x$ is at most $n$; for each guess, all steps can be computed in total time $\Oh(n)$, resulting in a run time of at most $\Oh(n^2)$ for the whole algorithm for some fixed $T$, and a total run time of $\Oh(n^2 \log(1/\eps))$.

\section{Conclusion}
\label{sec:concl}
In this work, we have managed to give a $(\sfrac43+\eps)$-approximate algorithm for the problem \PCmaxpm with run time in $\Oh(n^2 \log(1/\eps))$.
For any $\eps < \sfrac16$, this improves upon the previously best known approximation factor of $\sfrac32$ of a polynomial time algorithm by Deppert and Jansen~\cite{DJ19} for this problem.
Deppert and Jansen~\cite{DJ19} manage to get rid of the $\eps$-error on the solution by introducing an idea for the binary search called \enquote{Class Jumping}, by which they manage to bound the number of guesses needed by the binary search to find the actual optimal makespan $\OPT$ in terms of $\Oh(\log(c m)) \in \Oh(\log(n))$.
Employing this or a similar idea to our approach, one could improve the approximation factor to $\sfrac43$, with the run time likely being in $\Oh(n^2 \log(n))$.
If this can be done for our approach however is an open question.
Of even greater interest would be to find an (E)PTAS for \PCmaxpm, which is known for all the important variations of this problem, but not \PCmaxpm itself.
We hope our meticulous structural analysis of this problems provides approaches for future research in this direction.

\bibliographystyle{plain}
\bibliography{Bibliography}

\begin{thebibliography}{10}

\bibitem{DJ19}
Max~A. Deppert and Klaus Jansen.
\newblock Near-linear approximation algorithms for scheduling problems with
  batch setup times.
\newblock In {\em Proc. {SPAA} 2019}, pages 155--164. {ACM}, 2019.

\bibitem{F56}
L.~R. Ford and D.~R. Fulkerson.
\newblock Maximal flow through a network.
\newblock {\em Can. J. Math.}, 8:399–404, 1956.

\bibitem{graham79}
R.~L. Graham, E.~L. Lawler, J.~K. Lenstra, and A.~H.~G. Rinnooy~Kan.
\newblock Optimization and approximation in deterministic sequencing and
  scheduling: a survey.
\newblock {\em Ann. Discrete Math.}, 5:287--326, 1979.

\bibitem{HOR13}
Raymond Hemmecke, Shmuel Onn, and Lyubov Romanchuk.
\newblock n-fold integer programming in cubic time.
\newblock {\em Math. Program.}, 137(1-2):325--341, 2013.

\bibitem{HS87}
Dorit~S. Hochbaum and David~B. Shmoys.
\newblock Using dual approximation algorithms for scheduling problems
  theoretical and practical results.
\newblock {\em J. {ACM}}, 34(1):144--162, 1987.

\bibitem{JKMR19}
Klaus Jansen, Kim{-}Manuel Klein, Marten Maack, and Malin Rau.
\newblock Empowering the configuration-ip: new {PTAS} results for scheduling
  with setup times.
\newblock {\em Math. Program.}, 195(1):367--401, 2022.

\bibitem{JL16}
Klaus Jansen and Felix Land.
\newblock Non-preemptive scheduling with setup times: {A} {PTAS}.
\newblock In {\em Proc. {EUROPAR} 2016}, volume 9833 of {\em LNCS}, pages
  159--170. Springer, 2016.

\bibitem{MMMR15}
Alexander M{\"{a}}cker, Manuel Malatyali, Friedhelm~Meyer auf~der Heide, and
  S{\"{o}}ren Riechers.
\newblock Non-preemptive scheduling on machines with setup times.
\newblock In {\em Proc. {WADS} 2015}, volume 9214 of {\em LNCS}, pages
  542--553. Springer, 2015.

\bibitem{M59}
Robert McNaughton.
\newblock Scheduling with deadlines and loss functions.
\newblock {\em Manage. Sci.}, 6(1):1–12, October 1959.

\bibitem{MP89}
Clyde~L. Monma and Chris~N. Potts.
\newblock On the complexity of scheduling with batch setup times.
\newblock {\em Oper. Res.}, 37(5):798--804, 1989.

\bibitem{MP93}
Clyde~L. Monma and Chris~N. Potts.
\newblock Analysis of heuristics for preemptive parallel machine scheduling
  with batch setup times.
\newblock {\em Oper. Res.}, 41(5):981--993, 1993.

\bibitem{SW99}
Petra Schuurman and Gerhard~J. Woeginger.
\newblock Preemptive scheduling with job-dependent setup times.
\newblock In {\em Proc. {SODA} 1999}, pages 759--767. {ACM/SIAM}, 1999.

\bibitem{XZ00}
Wenxun Xing and Jiawei Zhang.
\newblock Parallel machine scheduling with splitting jobs.
\newblock {\em Discret. Appl. Math.}, 103(1-3):259--269, 2000.

\end{thebibliography}
\end{document}